\documentclass[eqsecnum, floatfix, superscriptaddress, preprint, aps, nofootinbib]{revtex4}

\usepackage{amsmath,amssymb,amsthm}
\usepackage{xspace}
\usepackage{graphicx} %for graphics
\usepackage{latexsym} %for \longrightarrow
\usepackage[mathscr]{eucal} %for \mathscr{}
\usepackage{bm}
\usepackage{dcolumn}
\usepackage{array}
\usepackage[usenames,dvipsnames]{color}

\begin{document}
\title{Recent advances in contextuality tests}

\author{Jayne \surname{Thompson}}
\email{thompson.jayne2@gmail.com}
\affiliation{Centre for Quantum Technologies, National University of Singapore, 3 Science Drive 2, 117543 Singapore, Singapore}

\author{Pawe{\l} \surname{Kurzy\'nski}}
\affiliation{Centre for Quantum Technologies, National University of Singapore, 3 Science Drive 2, 117543 Singapore, Singapore}
\affiliation{Faculty of Physics, Adam Mickiewicz University, Umultowska 85, 61-614 Pozna\'{n}, Poland}

\author{Su-Yong \surname{Lee}}
\affiliation{Centre for Quantum Technologies, National University of Singapore, 3 Science Drive 2, 117543 Singapore, Singapore}

\author{Akihito \surname{Soeda}}
\affiliation{Centre for Quantum Technologies, National University of Singapore, 3 Science Drive 2, 117543 Singapore, Singapore}

\author{Dagomir \surname{Kaszlikowski}}
\email{phykd@nus.edu.sg}
\affiliation{Centre for Quantum Technologies, National University of Singapore, 3 Science Drive 2, 117543 Singapore, Singapore}
\affiliation{Department of Physics, National University of Singapore, 2 Science Drive 3, 117542 Singapore, Singapore}

\pacs{03.67.-a, 03.65.Ta, 03.65.Ud}
\begin{abstract}
Our everyday experiences support the hypothesis that physical systems exist independently of the act of observation. Concordant theories are characterized by the objective realism assumption whereby the act of measurement simply reveals preexisting well defined elements of reality. In stark contrast quantum mechanics portrays a world in which reality loses its objectivity and is in fact created by observation. Quantum contextuality as first discovered by Bell [Rev. Mod. Phys. {\bf 38}, 447 (1966)] and Kochen-Specker [J. Math. Mech. {\bf 17}, 59 (1967)] captures aspects of this philosophical clash between classical and quantum descriptions of the world. Here we briefly summarize some of the more recent advances in the field of quantum contextuality. We approach quantum contextuality through its close relation to Bell type nonlocal scenarios and highlight some of the rapidly developing tests and experimental implementations.

\end{abstract}

\maketitle

\section{Introduction}

The aim of this paper is to survey some of the rapidly evolving ideas in the field of quantum contextuality. It can be argued that this field was in suspended animation during the intervening period between its conception in seminal papers by Bell and Kochen and Specker in 1967 \cite{Bell,KS} and the 2007 landmark discovery of a simple and experimentally verifiable inequality by Klyachko et al.  \cite{Klyachko}. Post 2007 there has been steady development of quantum contextuality. However in the view of the authors of this paper, the field is currently in a nascent interlude and still has undeveloped potential capable of creating an explosion of mainstream interest. With this in mind we would like to give a succinct summary of the progress on quantum contextuality at this junction. We also present a uniform approach to quantum contextuality and what is commonly referred to as  quantum nonlocality\footnote{The term quantum non-locality is imprecise causing many unnecessarily heated debates. It is obvious that quantum mechanics is local \cite{Haag}. We use this term to express the fact that quantum mechanics cannot be described by local realism. } which we believe delivers a stable platform for developing new results.

Through this platform we hope to find applications for quantum contextuality and thereby move it out of the realm of foundations of quantum mechanics and into mainstream quantum information science. Quantum  nonlocality has two undeniably important practical applications: device independent quantum key distribution \cite{Ekert, SCARANI} and private amplification of randomness \cite{ACIN}. Because quantum nonlocality is a special case of  contextuality it should also be possible to find practical applications for quantum contextuality. Once this happens we will witness a significant amplification of research in this area.

\subsection{Pre-history}

Quantum mechanics has always struggled with interpretational issues to the extent where its founding fathers famously contested the theoretical foundations of the new field. This struggle resulted in a seminal paper by Einstein, Podolsky and Rosen (EPR) in 1935 \cite{EPR} which introduced a philosophical manifesto called  local realism. From EPR's perspective local realism is a very fundamental property of Nature that underpins every physical theory.  The local realism hypothesis purports that observable properties of physical systems should exist objectively (i.e. properties of states exist independently of the observer) and concurrently the outcome of measuring any physical property can not be influenced by space like separated events. It is worth pointing out that all relativistic theories of matter obey local realism. However, as demonstrated experimentally by Aspect et al. in 1982 \cite{Aspect}, quantum mechanics is not locally realistic.

%The local realism hypothesis purports that observable properties of physical systems should exist objectively, i.e. independently from any potential observer, furthermore observation of one of these physical properties should not instantaneously influence the other properties, which are spatially separated in the sense of general relativity theory.

\subsection{Hidden variables}

EPR effectively started a programme of constructing hidden variable theories \cite{PM3}. Here the unilateral assumption is that the outcomes of measurements of physical properties reveal the underlying well defined properties of the measured system. For instance, if one measures the spin of a quantum particle along direction ``$x$" and obtains value ``$+1$", (the spin points up along the ``$x$" direction)  a hidden variable theory would argue that this property of the spin (the property of being up along the ``$x$" direction) was revealed by the act of measurement. In contrast, quantum mechanics tells us that this property was created by the act of measurement \cite{Peresbook}. A general statement that the outcomes of experiments exist objectively is of no use if it cannot be empirically verified. Adjudicating between quantum mechanics and hidden variables requires both theories to be reformulated in terms of experimentally testable predictions.

In this paper we are specifically interested in a special class of hidden variable theories called  non-contextual hidden variables (NCHV). NCHV theories can be simply characterized in the following way. Let us assume that we can measure some properties of a given physical systems denoted as $A_1, A_2,A_3,\dots, A_N$, where each measurable property $A_i$ can yield outcomes $\{a_i\}$. The noncontextual assumption states that the outcome of measuring $A_i$ is independent of which other members of the set $\{A_1, A_2,A_3,\dots, A_N\}$ are coincidentally measured. It is important to note here that each outcome value $a_i$ can appear inherently randomly (determinism being a special case). The description noncontextual means that the outcomes of measurements do not depend on the context in which they are measured, i.e., the outcome of $A_i$ is independent of whether it was measured in the context of $A_j$ or $A_k$.

\subsection{Bell-Kochen-Specker theorem}\label{sec:BellKochenSpecker}

The original question of contextuality of quantum theory is based on elementary events that are represented by projectors in Hilbert space. Two projectors are orthogonal, if the corresponding events are exclusive. The impossibility of outcome assignment, i.e., the impossibility of NCHV preassigning outcomes to events is formalized by the Bell-Kochen-Specker-theorem \cite{Bell, KS,Garcia-Alcaine, Peres} (BKS theorem):

\vspace{3mm}

\emph{In dimension greater than two; there exist projectors $\Pi_{i}$ such that it is not always possible to assign outcomes $0 , 1$ to all projectors in such a way that}
\begin{enumerate}
\item[(i)]\emph{For any complete orthogonal basis of projectors $\{\Pi_i\}$ the assignment satisfies $\Sigma_{i} \Pi_{i} =  1$ (completeness of the basis)}
\item[(ii)]\emph{If a projector, $\Pi_{i}$, belongs to multiple complete bases then it is consistently assigned the same value in all bases (noncontextual condition).}
\end{enumerate}
%\end{theorem}
The formulation of the BKS theorem with projectors can be considered as the most fundamental, since projectors correspond to elementary events that can either happen or not. However, the problem of contextuality was also defined for observables that are linear combinations of projectors. An elegant simple example of contextuality with $\pm 1$ observables is given in-terms of the Mermin-Peres square \cite{Peres, Mermin1, Mermin2}.

Consider a bipartite system of two spin $1/2$ particles. Take the square array:
\begin{align}\label{tab:merminperes}
\sigma_z \otimes \mathbb{I} && \mathbb{I} \otimes \sigma_z && \sigma_z\otimes \sigma_z \nonumber \\
\mathbb{I} \otimes \sigma_x && \sigma_x \otimes \mathbb{I} && \sigma_x\otimes \sigma_x \nonumber \\
\sigma_z \otimes \sigma_x  && \sigma_x \otimes \sigma_z && \sigma_y \otimes \sigma_y,
\end{align}

\noindent where $\sigma_x, \sigma_y, \sigma_z$ are Pauli matrices and $\mathbb{I}$ is the identity matrix and the operators have been arranged so that columns and rows form mutually commuting triads. Note that the product of the triad of operators in the last column is $-\mathbb{I}\otimes \mathbb{I}$, while in every other row and column the product is $\mathbb{I}\otimes \mathbb{I}$.  The Mermin-Peres square is a state-independent proof of the BKS theorem because a value assignment of $\pm 1$ to each operator in the above array can not reproduce this algebra \cite{Peres, Mermin2, Aravind}. This became an enduring foundational arguments for quantum contextuality.

\subsection{Mathematical roots of contextuality}

Noncontextual hidden variables exist if and only if the outcomes of all observable properties of a given physical system can be described by a joint probability distribution \cite{Fine}. Let us elaborate on this statement as it plays a central role in our review and provides a mathematically rigorous approach to NCHV theories. We will also focus on this topic in the next section.

Consider a set of observables $\{A_1, A_2, A_3,\dots\}$ that represent some physical properties of a quantum system. In general the corresponding operators, which we denote by the same letters $A_1, A_2, A_3,\dots$ do not all commute. According to quantum mechanics it is possible to simultaneously measure a set of observables if and only if the associated operators are mutually commutative. To make this statement clear we consider three observables $A_1,A_2,A_3$ such that $[A_1,A_2]=0, [A_1,A_3]=0$ but $[A_2,A_3]\neq 0$. Quantum mechanics gives us two experimentally verifiable probability distributions $p(a_1,a_2)$ and $p(a_1,a_3)$ and nothing beyond this. However, NCHV would assume the existence of a joint probability distribution for all observables, i.e., $p(a_1,a_2,a_3)$. Experimental veracity of this assumption would require that both the experimentally accessible distributions $p(a_1, a_2)$ and $p(a_1,a_3)$ can be consistently recovered as marginals from some well defined joint probability distribution:  $p(a_1,a_2)=\sum_{a_3}p(a_1,a_2,a_3)$ and $p(a_1,a_3)=\sum_{a_2}p(a_1,a_2,a_3)$.

 We illustrate these concepts via a spinless quantum  particle in a one-dimensional space. The experimentally measurable properties are: momentum and position. We denote the respective outcomes $p$ and $x$. The noncontextual hidden variable description presupposes a joint probability distribution $p(x,p)$. We can ask whether this joint provability distribution  $p(x,p)$  is supported by quantum mechanics, which simply means that the marginal probability distributions $p(x)$ and $p(p)$ must be recovered from $p(x,p)$ by appropriate integration. Here noncommutativity of the two operators guarantees a  joint probability distribution through $p(x,p)=p(x)p(p)$. Discrepancies between noncontextual hidden variable theories and quantum mechanics only arise when both commuting and noncommuting observables are present.

\subsection{Contextuality as a generalization of nonlocality}

We note that local realism is a special case of noncontextual hidden variables. Here spacelike separated measurements naturally correspond to commutative operators. Given two spacelike separated observers: Alice and Bob who have adjustable measurement settings, if Bob decides to measure an observable $B_1$ then each of Alice's measurement settings furnishes a context for $B_1$. We say the theory is noncontextual if Bob's measurement outcome is independent of Alice's settings. Due to the spacelike separation between Bob's measurement and Alice's choice of settings we refer to this scenario as local realism.  On the other hand, the general formulation of contextuality allows measurements to be performed on a single system. Therefore, the compatibility of measurements does not have to follow from the spacelike separation of observers.

It is an ongoing challenge to generalize all facets of nonlocality to contextual scenarios. In particular, despite the fact that nonlocality has found many computational applications, it is still unknown if general contextuality can be considered as a meaningful computational resource. We find the question whether single system contextuality, where nonlocal correlations and communication are no longer useful resources, can be used to outperform classical algorithms fascinating.

\subsection{Topics in this paper}

In Section \ref{sec:Anoverview} we approach contextuality by identifying it with the nonexistence of a joint probability distribution. In the process we highlight the close relationship between nonlocal and general noncontextual scenarios and some of the recent discoveries in the field of contextuality. In Section \ref{sec:contextualityinequalities} we review some prominent test of the contextuality of quantum theory. We naturally bipartition these tests into state-dependent and state-independent tests of contextuality. In Section \ref{sec:differentapproaches} we present some alternative approaches and properties of contextuality that are based on previous studies on nonlocality. Finally, in Section \ref{sec:experiment} we discuss some recent experiments that verified both, state-dependent and state-independent contextuality.

\section{ An overview of contextuality}
\label{sec:Anoverview}

We begin by identifying some key concepts which can be used to approach the discussion of contextual inequalities and experiments in the proceeding sections. We discuss how contextuality arises in systems possessing some measurable properties, $\{A_1, A_2, \dots A_N \}$,  whose measurement statistics can not be described by a joint probability distribution; an approach which is also used to study Bell inequalities and the problem of nonlocality. We will draw on this common framework to establish contextuality and its relation to the pre-established seminal work on non-existence of joint probability distributions in nonlocal scenarios.

We continue the perspective that contextuality is a more general feature than nonlocality and that many nonlocal properties should be somehow manifested in a more general form when considered in contextual scenarios. Application of  Bell-like inequalities to general contextuality is a new field of research and some properties of new contextual inequalities have only recently been discovered.

\subsection{Joint probability distributions}

Consider the following problem. An experiment returns a large amount of data corresponding to outcomes of measurements ${\cal M}=\{A_1,A_2,\dots, A_N\}$. The data was collected in many repetitions of the experiment. It is assumed that each measurement $A_i$ refers to some physical property of the system that is being measured and that its outcomes $a_i$ are discrete. The data allows you to estimate probabilities of events like $p(A_i=a_i)$, or $p(A_i=a_i,A_j=a_j)$. However, imagine that due to some reason in a single run of the experiment only special subsets $\{A_i,A_j,\dots\} \subset {\cal M}$ were measured, hence you are limited to estimate probabilities $p(a_i,a_j,\dots)$ corresponding to these subsets. Here, we use the simplified notation $p(A_i=a_i)\equiv p(a_i)$. Your task is to determine if a joint probability distribution $p(a_1,a_2,\dots,a_N)$ that can give a rise to all measurable marginal probabilities
\begin{equation}
p(a_i,a_j,\dots)=\sum_{x,y,\dots \neq i,j,\dots} p(a_1,a_2,\dots,a_x,\dots,a_y,\dots,a_N)
\end{equation}
can in principle exist.

What is implied by the existence of a joint probability distribution? Any such probability distribution has to be supported on a set of events. To characterize an event we preassigned outcomes of all measurable quantities; in this case events are of the form: $A_1=a_1,A_2=a_2,\dots,A_N=a_N$. Therefore the existence of a joint probability distribution implies that it is possible to preassign outcomes to measurements before the actual measurement happens. In this scenario the measurement outcome of every property existed independently of the act of observation.
Furthermore within the set of co-measurable properties $\{A_i, A_j, \dots \}$, the outcome of measuring $A_i$ is independent of which other members of the set are coincidentally measured. This is often called the {\it objective realism} hypothesis. It is a hallmark of systems governed by the laws of classical physics. In quantum mechanics this possibility is often associated with some non accessible hidden variables, that together with a state vector $|\psi\rangle$ would be able to  determine outcomes of all measurements on the system.

On the other hand, if the outcomes of measurements do not exist before the measurement is performed (to which one often refers as to lack of {\it realism}), or if the outcomes depend on what other measurements are done at the same time ({\it contextuality}), a joint probability distribution does not exists. This fact was first observed for nonlocal scenarios in Ref. \cite{Fine}, and was later extended to contextual scenarios in Ref. \cite{LSW}.

\subsection{Complementarity and compatibility relations}\label{sec:complementarity}

We have shown that the problem of the existence of a joint probability distribution arises due to the fact that not all measurements can be performed in a single experiment. Following quantum mechanics, we refer to this property as to {\it complementarity}. On the other hand, if two measurements can be performed in the same experiment we call them {\it compatible}. It is therefore natural to ask what kind of complementarity and compatibility relations can give rise to the lack of a joint probability distribution. This problem was studied in Ref. \cite{Ramanathan}. Below we summarize the main results.

First we introduce graphs representing complementarity relation, where each vertex corresponds to some measurement and two vertices are connected by an edge if the two measurements are compatible.
\begin{figure}[hbt!]
\begin{center}
\includegraphics[width=0.6\columnwidth]{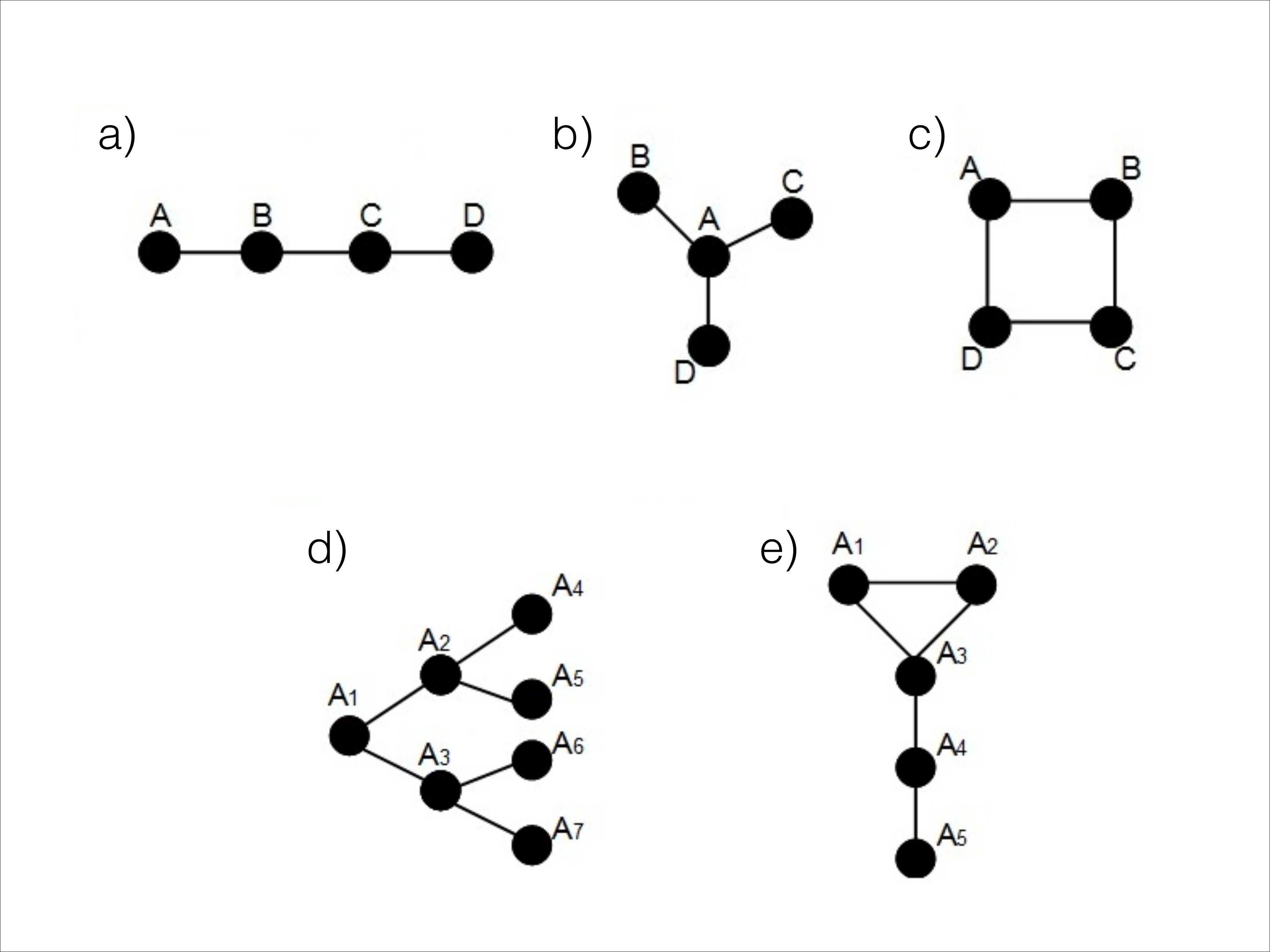}
\caption{\small Examples of complementarity graphs}
\label{fig:jpd1}
\end{center}
\end{figure}
An example of such graph is presented in Figure \ref{fig:jpd1}. Every clique in the graph, i.e., a subgraph in which all vertices are mutually connected, forms a set of measurements for which joint probability distributions can be experimentally estimated. For example, the graph in Figure \ref{fig:jpd1} a) consists of four measurements: A, B, C and D. Probabilities that can be measured are: $p(a,b)$, $p(b,c)$, $p(c,d)$, $p(a)$, $p(b)$, $p(c)$ and $p(d)$. Can one find a joint probability distribution $p(a,b,c,d)$ capable of reproducing measurable marginals?

In case of example a) such probability distribution exists. It can be constructed in the following way
\begin{equation}
p(a,b,c,d)=\frac{p(a,b)p(b,c)p(c,d)}{p(b)p(c)}.
\end{equation}
Note, that although this construction may seem artificial, due to complementarity relation one is not able to disproof its validity. Moreover, the above construction recovers all measurable marginals. For example
\begin{eqnarray}
p(b,c) &=& \sum_{a,d}p(a,b,c,d)=\sum_a\left(\sum_d \frac{p(a,b)p(b,c)p(c,d)}{p(b)p(c)} \right) \nonumber \\ &=& \sum_a \frac{p(a,b)p(b,c)}{p(b)} = p(b,c).
\end{eqnarray}
The above follows from the fact that for all $x$ and $y$ one has $p(x)=\sum_y p(x,y)$ (we will come back to this problem later).

Joint probability distributions can be constructed in an analogous way for all complementarity graphs that do not contain cycles of size greater than three. 3-cycles are already cliques, therefore one can experimentally estimate a joint probability distribution for all three measurements in the cycle. The construction goes as follows. First, we multiply joint probability distributions for all cliques (3-cycles and two vertices connected by an edge) in the graph and next we divide by the joint probabilities of measurements that appear more than once. If a measurement, say A, appears $n$-times, one divides the product of probabilities of cliques by $p(a)^{n-1}$. For more details see \cite{Ramanathan}. For example, a joint probability distribution for the case b) (Figure \ref{fig:jpd1}) can be constructed as
\begin{equation}
p(a,b,c,d)=\frac{p(a,b)p(a,c)p(a,d)}{p(a)^2}.
\end{equation}

The marginal probabilities can be obtained by summing over all other measurements. However, summation must be performed in the specific order. We start summation from the measurements that in the construction appear only in one clique. After the summation, some terms cancel and we obtain a new distribution. Next, we sum over measurements that in the new distribution appear only once. We follow this recursive procedure as long as a desired marginal probability is obtained.

For graphs that contain cycles of size greater than three the above construction does not work. This is because of two reasons. First, $n$-cycles ($n>3$) are not cliques themselves, therefore one cannot experimentally evaluate probabilities for all measurements in the cycle. Moreover, in $n$-cycles every vertex appears in more than one clique and the above summation procedure cannot be performed. Therefore one arrives at the necessary condition for the lack of joint probability distribution. Namely, the lack of a joint probability distribution can be detected only in scenarios in which complementarity relation between measurements induces cycles of size greater than three. In the Figure \ref{fig:jpd1} only the case c) may not admit a joint probability distribution. Note, that this case corresponds to a well known Clauser-Horne-Shimony-Holt (CHSH) nonlocal scenario \cite{Clauser} in which four measurements are done on a bipartite system. Two complementary measurements (say A and C) are done on the first subsystem, whereas the other two (B and D) are done on the second subsystem. For a detailed study of probability distributions and inequalities based on $n$-cycles see Ref. \cite{ncycles}.

\subsection{Non-contextuality assumption}

Every Bell-like inequality relies on three assumptions: free will, realism and non-contextuality. The free will assumption, i.e., observers can freely choose which measurements to perform, is rarely questioned on the scientific grounds \cite{CK}. Moreover, the joint probability distribution approach does not require the discussion of the free will assumption. Therefore, we focus on the remaining two assumptions.

The notions of nonlocality and contextuality refer to our classical intuition. In classical physics values of physical parameters associated with outcomes $a_i$ exist independently of whether we decide to measure them or not. In this sense classical theories are often said to be realistic. Sometimes measurement outcomes of a  realistic theory cannot be described by a joint probability distribution. This happens when the outcome of measuring a property depends on what other measurements are performed at the same time. In general this dependence is called contextuality, since the outcome depends on the measurement context.  In the special case where the measurements are performed in different spatial locations the same dependence is known as nonlocality. Putting aside the free will assumption, present experiments cannot uniquely determine whether the universe lacks realism or is contextual in the above sense (or both). Nevertheless, it is commonly accepted to refer to quantum phenomena which exhibit these effects as to quantum contextuality and quantum nonlocality respectively.

The non-contextuality assumption states that measurement outcomes do not depend on what other measurements are performed at the same time.
As this is a key point we highlight it by using an explicit example. Consider a system with measurable properties ${\cal M}= \{A_1, A_2, \dots A_N\}$. Take all possible subsets comprised of jointly co-measurable properties. Let ${\it m}_k = \{A_i,A_j,\dots\}\subset {\cal M}$ be a representative subset; we say $m_k$ furnishes a measurement context for each of its elements $A_i,A_j,\dots$. The system is noncontextual if we can uniquely determine the measurement statics of a property $A_i$ by sampling its outcomes $a_i$ in measurement context $m_k$. In other words for every property $A_i \in {\cal M}$, the measurement outcomes of this property are the same irrespective of which measurement context is chosen.

In case of nonlocal Bell-type scenarios the assumption that measurement outcomes of Alice do not depend on spacelike separated measurement choices made by Bob is supported by the special theory of relativity which dictates no interaction can spread faster than the speed of light. It is therefore assumed that any potential interaction that could cause violation of some Bell inequality would have to obey special relativity, and hence it would have no effect if two measurements were spatially separated. On the other hand, in contextual scenarios in which measurements are performed on the same system one lacks additional support for non-contextuality coming from other physical theories. In this case non-contextuality is rather a reasonable assumption that is supported on our realistic everyday experience that values of physical properties are independent of what other properties are observed at the same time. This is because the measurement is only assumed to reveal the value of the measured property without any interference.

It has been demonstrated by Fine \cite{Fine} that the existence of such a joint probability distribution is necessary and sufficient conditions for constructing a non-contextual hidden variable theory capable of replicating the measurement statistics.

\subsection{Noncontextuality assumption in quantum theory}
\label{sec:noncontextualityinquantumtheory}

The smallest contextual quantum system has three levels and can be physically realized by a spin 1 system. Due to its relative simplicity, the spin 1 system is often used to study contextuality. Here it is possible to interpret measurable properties as projectors onto eigenstates $|S_{x}=0\rangle$ of the spin operators $S_x$ which measures spin relative to a direction $\vec{x}$. In quantum mechanics projectors onto the two spin 1 states $|S_{x}=0\rangle$ and $|S_{y}=0\rangle$ respectively, can be jointly measured if $\vec{x}$ and $\vec{y}$ are orthogonal. Physically, if the measurement projects the state of a spin 1 particle onto $|S_{x}=0\rangle$ one learns that {\it the direction of spin is orthogonal to $\vec{x}$}.

 A joint measurement of projectors onto $|S_{x}=0\rangle$ and $|S_{y}=0\rangle$ corresponds to a question: {\it  which of the two orthogonal directions $\vec{x}$ and $\vec{y}$ the spin direction is orthogonal to?}.

It follows that $|S_{x}=0\rangle$ can be measured together with $|S_{y}=0\rangle$ or $|S_{y'}=0\rangle$, where $\vec{y}$ and $\vec{y'}$ lie in the yz-plane. These two measurement possibilities create a context and one may ask how it is possible that the property: {\it the spin direction is orthogonal to $\vec{x}$}, depends on what other direction is investigated at the same time.

Obviously, there is no known physical principle stating that properties of the spin direction cannot change from one measurement context to another. Nevertheless, it would be very strange (according to our classical intuition) if they did.

\subsection{Bounds on contextuality}\label{sec:contextualrelationsbounds}

Studies of contextuality focus on theories in which the context dependence is not explicit in a sense that the outcome of measurements may depend on the context, but the probabilities are context independent. This means that for all measurements $A$, $B$ and $C$ one has $p(a)=\sum_b p(a,b)=\sum_c p(a,c)$, irrespective of whether $B$ and $C$ are compatible or complementary \cite{Gleason,CSW}. Quantum theory is one such theory. The violation of this assumption, i.e., $\sum_b p(a,b) \neq \sum_c p(a,c)$ leads to peculiar effects. For example, if $B$ and $C$ were performed on a different system than $A$, the violation of the above property would lead to a superluminal signaling. Therefore, in nonlocal scenarios the condition $\sum_b p(a,b)=\sum_c p(a,c)$ is known as the {\it no-signaling} condition. In general contextual scenarios we refer to this property as to {\it no-disturbance} \cite{Monogamy2}.

Contextual scenarios which probe the BKS theorem often assume that measurements correspond to events. In this case the compatibility relation takes the form of exclusivity. In quantum mechanics events are represented by projectors and exclusivity occurs if two projectors are orthogonal. The corresponding complementarity graphs represent the exclusivity relations (exclusivity graphs), where two vertices are connected if the corresponding events are exclusive. For some examples of how these graphs could look see Figure \ref{fig:jpd1}.

Let us now discuss experimental tests of inequalities based on exclusive events. First of all, for a pair of exclusive events A and B only one of three scenarios can happen in a laboratory: either the first event occurs ($A=1,B=0$), the second event occurs ($A=0,B=1$), or both events do not occur ($A=0,B=0$). Due to the no-disturbance condition $p(A=1,B=0)=p(a)$ and $p(A=0,B=1)=p(b)$, and exclusivity demands that $p(a)+p(b) \leq 1$, where $a$ (respectively $b$) is the event $A = 1$ (respectively $B = 1$). This has to hold for all exclusive events, therefore in all exclusivity graphs the sum of probabilities corresponding to connected vertices has to be bounded by one. Moreover, under the assumption that a set of mutually pairwise exclusive events is exclusive \cite{ICC}, the sum of probabilities in any clique has to be bounded by one too.

Despite the above restrictions on probabilities, one still has some freedom when assigning probabilities to vertices. The non-contextual bound in inequalities based on exclusive events is derived under the assumption that a joint probability distribution for all events exists. These inequalities take the form of the sums of individual probabilities of all events. The non-contextual bound was shown to be equal to the independence number $\alpha$ of the corresponding exclusivity graph, $G$, for the set of projectors $\{X\}$:
\begin{equation}\label{eq:generalforminequality}
\sum_{X\in V(G)} p(x) \leq \alpha
\end{equation}
where in the above $V(G)$ is the set of vertices belonging to the exclusivity graph $G$ and by $p(x)$ we mean the probability of getting outcome $X = 1$ when measuring projector $X$ \cite{CSW}. The independence number is the maximal number of mutually disjoined vertices in the graph, which in our case corresponds to the maximal number of events that can occur, under the assumption that whether event occurs or not does not depend on the measurement (realism).

The maximum value of the left hand side of Equation (\ref{eq:generalforminequality}) attainable by quantum observables was shown to correspond to the Lovasz $\vartheta$-function of $G$, which is another graph property obeying $\alpha\leq\vartheta$ \cite{CSW}. The Lovasz $\vartheta$-function is based on the orthogonal representation of graphs and is defined as
\begin{equation}
\vartheta (G)=\max_{|\psi\rangle, V} \sum_{|v_i\rangle\in V} |\langle v_i|\psi\rangle|^2,
\end{equation}
where $|\psi\rangle$ is a unit vector and $V$ is a set of unit vectors $|v_i\rangle$ corresponding to vertices of the graph having the property, that the two vectors are orthogonal if the corresponding vertices are connected by an edge. The most interesting scenario is when the Lovasz $\vartheta$-function of $G$ is greater than independence number; under these circumstances Inequality (\ref{eq:generalforminequality}) can be used to demonstrate contextuality of quantum theory.

No-disturbance and exclusivity allows for even greater violation of the inequality. However, recent results suggests that Lovasz $\vartheta$-function, and hence quantum mechanics, allows for greatest possible violation of contextual inequalities that are compliant with the exclusivity principle which states that the sum of probabilities of pairwise exclusive events is bounded by one. If one applies this principle to a set of events corresponding to a single test, one may obtain even greater violation than the violation allowed in quantum theory. However, if one applies this principle to a set of events corresponding to multiple independent tests (global exclusivity), one may find that the extended structure of subsets of pairwise exclusive events can lead to stronger bounds on contextuality \cite{ICC,ICC2}.

\section{Tests of contextuality}
\label{sec:contextualityinequalities}

In simple words, a contextuality test examines whether expectation values of a number of observables cannot be described via joint probability distributions. In cases considered in this work, this corresponds to a detection of the violation of certain inequalities that involve expectation values obtained in a laboratory. However because an expectation value is a linear combinations of probabilities, the inequalities can take an alternative form similar to Equation (\ref{eq:generalforminequality}).

The original BKS theorem does not depend on a state of the system. Quantum contextuality, as introduced by Kochen and Specker \cite{ KS} (and independently motivated by Bell \cite{Bell}), is a property of measurements, not a property of states.

 On the other hand quantum nonlocality is the lack of a joint probability distribution which can be constructed under the assumptions of local realism and noncontextuality. However, quantum nonlocality can only occur if the state of the measured system is entangled. This may seem to contradict our perspective that nonlocality belongs to the general framework of contextuality.

From the perspective of testing types of contextuality, nonlocality tests are state dependent because all permissable measurements on the quantum system must be implementable by local operations. This is a strong restriction on the set of measurements which can be performed.

In order to reveal the lack of a joint probability distribution for all states one needs to have enough freedom in choosing the measurement settings. The ability to preform a test which reveals contextuality for all quantum states (a state-independent test) is a property of a set of observables which have sufficiently exploited this freedom to enable detection of contextuality in the completely mixed state.

In multipartite scenarios this would involve global nonlocal measurements on all parts of the system. In case of the standard BKS scenario all measurements are performed on a single system, therefore in principle there are no restrictions on measurements.

In the following two sections we will be focussing on state-dependent and state-independent tests of contextuality.

State-independent tests are capable of revealing contextual behavior for any state of the quantum system. Explicitly, a state-independent test invokes a set of observables $\{A_1,A_2,\dots\, A_N\} $  such that for any state of the quantum system there is no joint probability distribution describing the outcome of measuring these observables on that state. While state-dependent tests typically use fewer observables to show that  no joint probability distribution can describe the measurement outcomes on certain nominated states of the quantum system.

In Subsection \ref{sec:statedepandindep} we will expand upon this topic by showing how to convert a specific state-independent tests into state-dependent tests. The general procedure is to restrict the collection of measurable observables.

\subsection{State-dependent contextual inequalities}

\label{sec:statedependent}
We would like to use the techniques from Subsection \ref{sec:contextualrelationsbounds} to derive state-dependent contextual inequalities. This is a critical step in the development of experimental tests of contextuality. We focus on developing ideas rather than listing the numerous state-dependent inequalities.

\subsubsection{Wright's inequality}

We begin with a game where systems exhibiting quantum contextuality can be interpreted as resource states which can boost the chances of winning \cite{AACB13}.

An observer has access to a classical system which can be prepared in any arbitrary mixture of five classical states, labeled $1,2,3,4,5$. They are tasked with ascertaining the answers to five yes-no questions. We write the first question in full as $A_1$: `Is the system in either state $1$ or $2$?", and in abbreviated form as
$A_1= ``1~or~2?"$. We use ``$A_1 = 1$" to denote an affirmative answer to this question and ``$A_1  = -1$" to denote a negative outcome. An observer's score is calculated using the sum of the probability of attaining a positive answer $p(A_i = 1)$, for each question $i =1,\dots, 5$ and can be interpreted as the number of questions answered affirmatively. To win it is sufficient to maximize your score. In this convention the remaining questions are $A_2= ``3~or~4?"$, $A_3= ``1~or~5?"$, $A_4= ``2~or~3?"$, and $A_5= ``4~or~5?"$. Because the outcomes $A_i=1$ and $A_{i\pm 1 \, {\rm mod }\, 5}=1 $ are mutually exclusive at most $2$  questions can be answered affirmatively and
\begin{eqnarray}\label{eq:wright}
\sum^{5}_{i=1}p(A_i =  1)\leq 2.
\end{eqnarray}
This is known as Wright's inequality \cite{Wright}. If the classical system is prepared in a random mixture of all states then the probability an affirmative answer is given to any question is $2/5$, which saturates the above inequality.

However, when the observer has access to quantum systems with dimension greater than $2$; quantum contextuality is linked to higher scores \cite{AACB13}. We prepare the system in a pure state
\begin{equation}\label{eq:spinupzstate}
\langle \psi | = (0,0,1),
\end{equation}
and translate our questions into dichotomic, $\{+1, -1\}$-observables of the form
\begin{equation}\label{eq:KCBSobservables}
A_i =  2|v_i\rangle \langle v_i | -1.
\end{equation}
Even when we fulfill the exclusivity condition, which occurs when the projectors $\Pi_i = |v_i\rangle \langle v_i|$ and $\Pi_{i\pm 1 \,{\rm mod} \, 5}$ are orthogonal, the maximum attainable value of the left hand side of Equation (\ref{eq:wright}) is
\begin{equation}\label{eq:KCBSviolation}
\sum^{5}_{i=1}|\langle v_i |\psi\rangle|^2  = \sqrt{5} \approx 2.236
\end{equation}
This scenario would correspond to optimal violation of Wright's inequality (\ref{eq:wright}) and a higher score in the game \cite{AACB13}. It is physically realizable when:

\begin{eqnarray}\label{eq:KCBSdirections}
&&\langle v_1| = N(1,0,\sqrt{c})\nonumber\\
&& \langle v_{2,5}| = N(- c, \pm s, \sqrt{c})\nonumber\\
&&\langle v_{3,4}| = N(C, \pm S, \sqrt{c})
\end{eqnarray}
for $c = {\rm cos} (\pi/5)$ and $s = {\rm sin}(\pi/5)$ while $C = {\rm cos} (2\pi/5)$ and $S = {\rm sin}(2\pi/5)$. The normalization constant $N = \sqrt{1+ c}$.

\subsubsection{KCBS inequality}

Wright's inequality is intimately related to the simplest state-dependent contextual inequality which is commonly referred to as the Klyachko-Can-Binicioglu-Shumovsky-inequality (KCBS inequality) \cite{Klyachko}.

Given a spin-$1$ system, in a sufficiently pure state, the KCBS inequality utilizes five observables whose measurement statics can not be replicated by a joint probability distribution or hidden variable model to demonstrate quantum contextuality. Here we consider the simplest case of a pure state system as it furnishes the strongest resemblance between the KCBS inequality and Wright's inequality.

\begin{figure}[hbt!]
  \begin{center}
 \includegraphics[width=0.6\columnwidth]{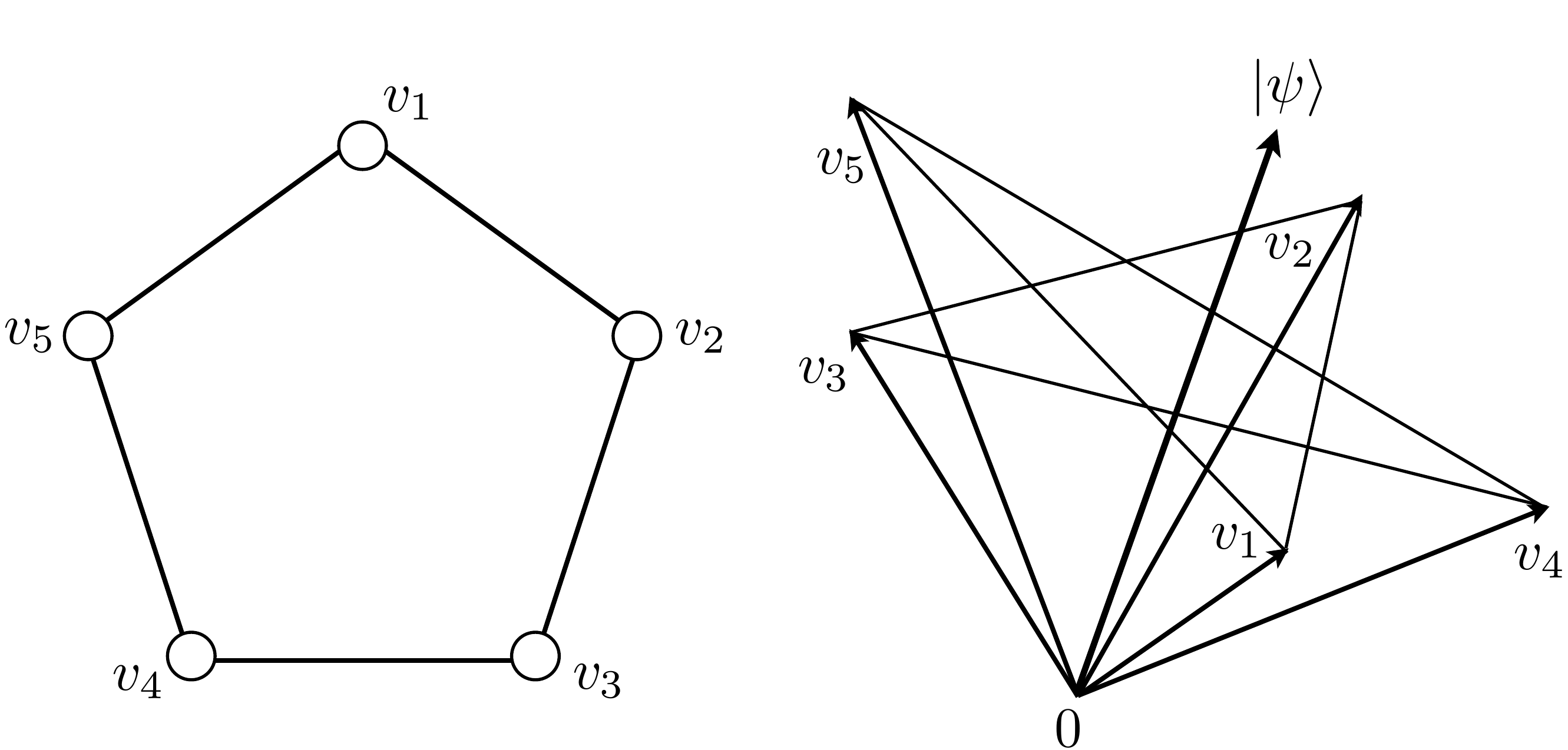}
  \end{center}  \caption{\small The 5-cycle graph describes the orthogonality (complementarity) relations between the five projectors from Equation \ref{eq:klyachkoprojectorineq}; where the 3-dimensional vectors ${\it v}_i$ are directed towards the five points of a star. Each projector is represented by a vertex of the 5-cycle. Vertices corresponding to compatible projectors are adjacent. Note this 5-cycle graph is isomorphic to the exclusivity graph where each vertex corresponds to an event $\Pi_i = |v_i\rangle \langle v_i| = 1$ for $i = 1,\dots, 5$ and exclusive events are linked by an edge. This implies the graph independence number of the 5-cycle is also the bound in Equation (\ref{eq:klyachkoprojectorineq}).  For more detail see Subsections  \ref{sec:complementarity} and \ref{sec:contextualrelationsbounds}. The vector $|\psi\rangle$ describes the relative orientation (with respect to the 5 projection eigenspaces) of the state optimally violating Inequalities (\ref{eq:klyachkoprojectorineq}) and (\ref{eq:klyachkoccorrelationineq}) }
  \label{fig:klyachko}
\end{figure}

 The five observables are projective measurements, $\Pi_i = |v_i\rangle \langle v_i|$ for $i =1,\dots 5$,  on the pure state. Pairs of projective measurement are considered to be compatible if and only if they are orthogonal\footnote{This follows directly from the discussion in Subsection \ref{sec:noncontextualityinquantumtheory}.}. Following the quantum version of Wright's inequality we choose consecutive observables $\Pi_i$ and $\Pi_{i\pm 1 \,{\rm mod} \, 5}$  to be compatible; these complementarity relations can be summarize by the $5$-cycle graph in Figure \ref{fig:klyachko}. This means we can construct well defined marginal probability distributions which predict the outcomes of measuring $\Pi_i$ and $\Pi_{i + 1 \, {\rm mod}\, 5}$ coincidentally.

One realization of these compatibility relations is when the system, $|\psi\rangle$, is given by Equation (\ref{eq:spinupzstate}) and the five projective measurements are onto the subspaces defined in Equation (\ref{eq:KCBSdirections}).

Because the events $\Pi_i = 1$ and $\Pi_{i\pm 1 \, {\rm mod}\, 5} = 1$  are exclusive we know that for the five projectors, which are identified with vertices of  Figure \ref{fig:klyachko}, at most two projectors can be simultaneously assigned value $1$. It is possible to derive this from the graph independence number of Figure \ref{fig:klyachko}. This implies that for non-contextual realistic theories:

\begin{equation}
\sum^5_{i =1} |\langle v_i|\psi\rangle|^2 \le 2.
\label{eq:klyachkoprojectorineq}
\end{equation}

Equation (\ref{eq:KCBSviolation}) demonstrates that this inequality is violated by  the pure state $\langle \psi | $. We pause to remark that the above inequality could be directly obtained from using Figure \ref{fig:klyachko} in conjunction with Equation (\ref{eq:generalforminequality}).

The KCBS inequality is conventionally reinterpreted as a statement about the correlations of the $\{ 1,-1\}$-dichotomic observables given in Equation (\ref{eq:KCBSobservables}). By simple substitution in Equation (\ref{eq:klyachkoprojectorineq}) it is possible to derive the conventional form of the KCBS inequality:
\begin{equation}
\langle A_1A_2\rangle +\langle A_2A_3\rangle  +\langle A_3A_4\rangle +\langle A_4A_5\rangle +\langle A_5A_1\rangle  \ge -3.
\label{eq:klyachkoccorrelationineq}
\end{equation}

Klyachko et al. discovered quantum mechanics predicted violation of the KCBS inequality by $5 -4\sqrt{5}$ when measurements are made on the optimal state $\psi$ depicted at the center of the pentagram in Figure \ref{fig:klyachko}; the KCBS inequality will be violated by a small solid angle of states surrounding $\psi$ \cite{Klyachko}.

 Lets take a closer look at the 5-cycle graph in Figure \ref{fig:klyachko}. In Subsection \ref{sec:complementarity} we established a necessary condition for being unable to construct a joint probability distribution; we must have a collection of $(n\ge 4)$-observables whose complementarity relation form a $n$-cycle graph. Furthermore we identified the Clauser-Horne-Shimony-Holt-inequality (CHSH inequality) with the case of 4-observables.

 In this framework the CHSH and KCBS inequality which correspond to $4$-cycle and $5$-cycle graphs respectively, are the simplest tests of quantum nonlocality and quantum contextuality in the sense that they use the minimal number of observables necessary for no joint probability distribution to exist and the smallest possible systems.

\subsubsection{9-projector contextuality for all states except the maximally mixed one}

\begin{figure}[ht!]
  \begin{center}
 \includegraphics[width=0.4\columnwidth]{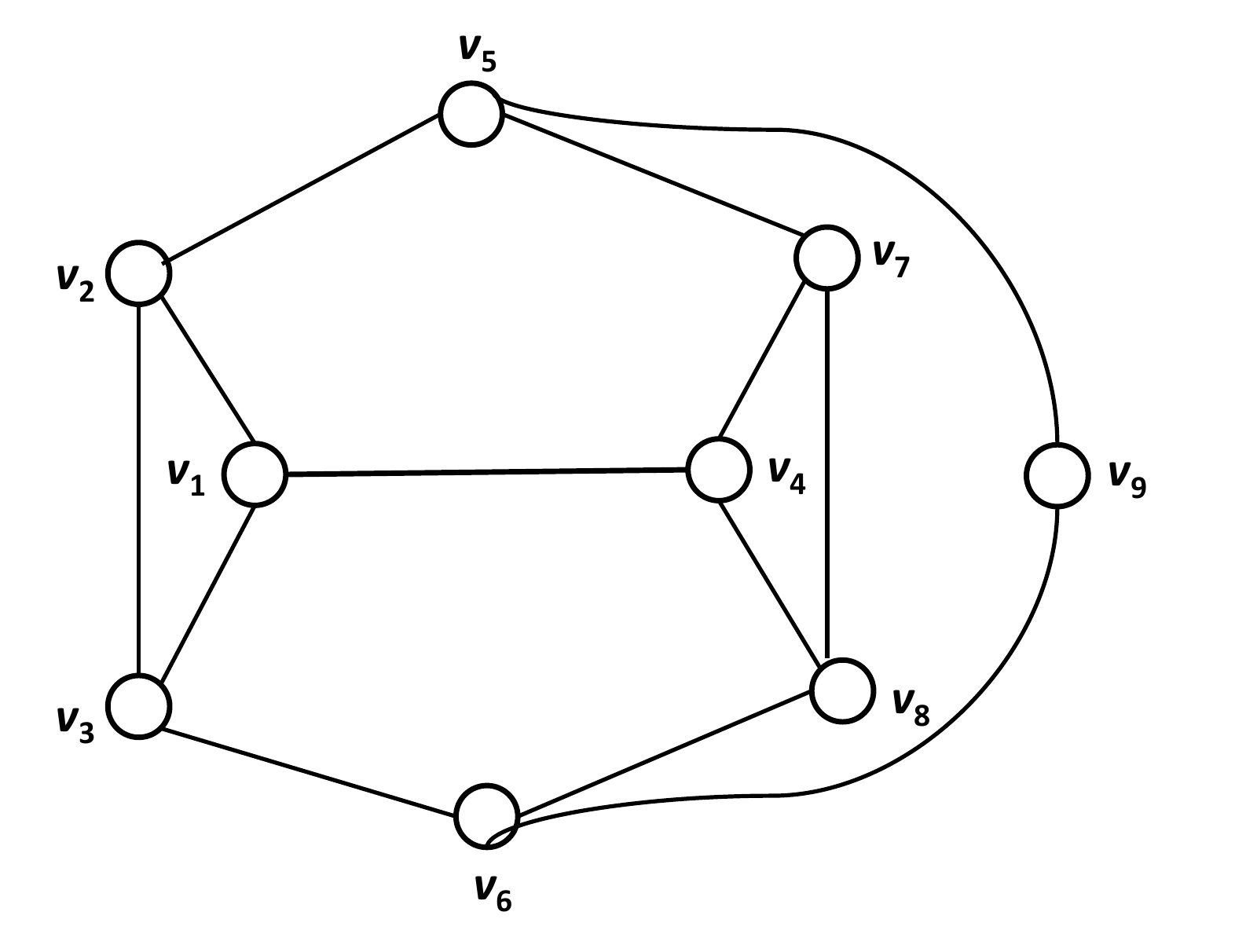}
  \end{center}
  \caption{\small The graph $G$ of the compatibility relations of the nine observables in Inequality \ref{dagandpawelcorrelationineqaulity}. Comparability of two observables $A_i$ and $A_j$ is furnished by orthogonality of the corresponding projectors $\Pi_i$ and $\Pi_j$.}
  \label{fig:dagandpawel}
\end{figure}

Contextual inequalities can often have complementarity relation graphs with more structure than the simple $n$-cycles we have presented so far in this paper, which can lead to violation of the inequality by a larger class of states. To illustrate these developments we introduce a nine-projector inequality presented in Ref. \cite{Kurzynski} which can detect contextuality for any quantum state of a spin-1 particle, except the maximally mixed state, provided we choose the measurement basis appropriately. We give the form of the nine projector inequality based on dichotomic $\pm 1$ observables $A_i = 2|v_i\rangle \langle v_i| -1$:

\begin{equation}\label{dagandpawelcorrelationineqaulity}
\sum_{(i,j) \in E(G)} \langle A_i A_j \rangle -\langle A_9\rangle  \ge -4.
\end{equation}
where $E(G)$ is the edge set of Figure \ref{fig:dagandpawel} which describes the complementarity relations of the nine observables. For an explicit realization of how these 9-projective measurements can be oriented in a 3-dimensional Hilbert space see Ref. \cite{Kurzynski}.

It can be shown that for any state, except the maximally mixed one,  one can always find a basis for projectors $\Pi_i=|v_i\rangle \langle v_i| $ such that both, the state and the matrix $\sum_{i=1,\dots 9} \Pi_i$, are diagonal in this basis. Moreover, in this case the state will always violate Inequality (\ref{dagandpawelcorrelationineqaulity}) unless it is completely mixed. The completely mixed state saturates the inequality irrespective of the choice of basis. If the measurement basis is not individually tailored to each quantum state then numerical simulations, using QI Mathematica package \cite{Miszczak} and ten million randomly generated density matrices weighted by the Hilbert-Schmidt measure, have demonstrated that $ 49.98\% $ of quantum states violate Inequality (\ref{dagandpawelcorrelationineqaulity}) \cite{Thompson}.

An interpretation of this result can be given when there are two cooperating parties, Alice and Bob. Assume Alice has access to an ensemble of spin-1 particles from which she can prepare a density matrix $\rho$.  Bob receives the density matrix from Alice and executes the test. If Bob is completely ignorant of Alice's state preparation procedure then from his perspective the system is in a completely mixed state and he can at best saturate Inequality (\ref{dagandpawelcorrelationineqaulity}). However if Bob receives some information about the preparation process from Alice then he will be able to violate the inequality.

\subsubsection{State-dependent contextuality based on more complex events}

\begin{figure}[ht!]
  \begin{center}
 \includegraphics[width=0.6\columnwidth]{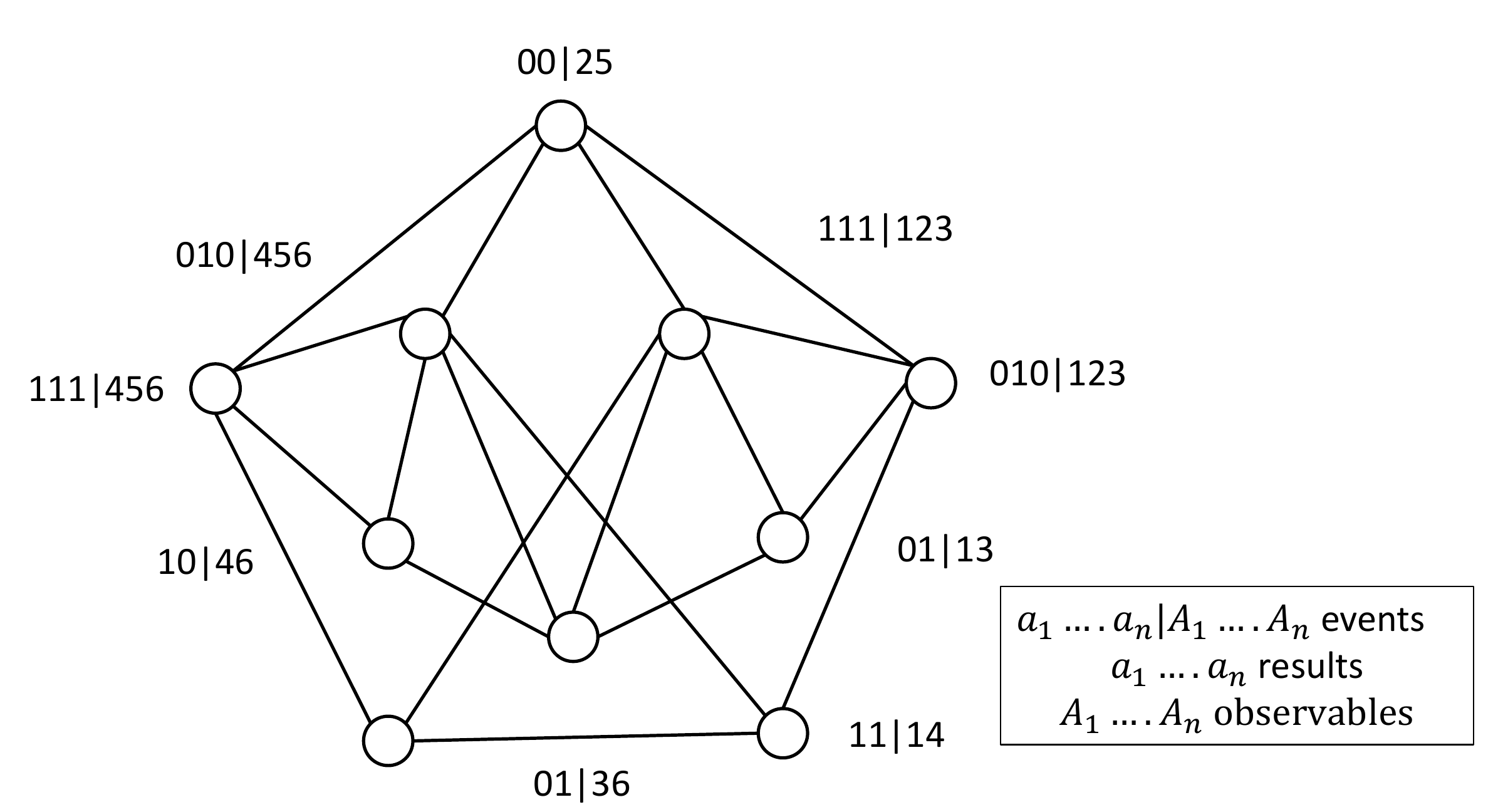}
  \end{center}
  \caption{\small The graph of exclusive events for the $6$ $\{1,0\}$-observables in Equation (\ref{eq:Cabelloexperiment}). Events correspond to assigning an outcome to a collection of compatible observables $A_1,\dots, A_n$. Two events are deemed exclusive if they can not both happen simultaneously. Each event is assigned to a vertex of the graph and exclusive events are confected by an edge. The upper bound on Inequality (\ref{eq:Cabelloexperiment}) is directly given by the independence number of this graph.}
  \label{fig:cabellorefpic1}
\end{figure}

To finish our discussion on state-dependent contextuality we highlight  an inequality  derived in exactly the same way as Wright's inequality in Equation (\ref{eq:wright}) \cite{ADLPBC12}. This  characterizes a different format for expressing contextual inequalities. Consider a collection of $6$ observables $A_1,\dots, A_6$. These observables have complementarity relations which permit only certain subsets, $A_1,\dots, A_n$, to be measured coincidentally. Parallelling the construction of Wrights inequality we identify events of the form $A_1 = a_1, \dots A_n = a_n$, where $a_i=0,1$, which we will abbreviate to $a_1,\dots, a_n|1,\dots, n$. Of particular interest are the events whose exclusivity relations are outlined in Figure \ref{fig:cabellorefpic1}.

When a joint probability distribution exists the maximum number of these events which can occur simultaneously is given by the independence number of the graph in Figure \ref{fig:cabellorefpic1} (see Subsection \ref{sec:contextualrelationsbounds}). Using the same logic as in Wright's inequality it follows that:

\begin{eqnarray}\label{eq:Cabelloexperiment}
&&p(010|123)+ p(111|123)+ p(01|13)  + p(00|14) + p(11|14) + p(00|25) + \nonumber \\ &&  p(01|36) + p(010|456) + p(111|456) + p(10|46)\leq 3,
\end{eqnarray}
where by $p(a_1,\dots , a_n|1,\dots, n)$ we simply mean the probability of the corresponding event occurring.
It was discovered in Ref.  \cite{ADLPBC12}  that the quantum violation of this inequality can be up to $3.5$.

This can be compared directly to Wright's inequality. The major distinction between these two scenarios is that the events in Equation (\ref{eq:Cabelloexperiment}) are more complicated; they require the outcomes of a collection of compatible observables to be specified. In contrast each event in Wright's inequality can be uniquely labeled by the outcome of a single observable.

\subsection{State-independent contextuality}

The Bell-Kochen-Specker theorem proves that contextuality arises from the algebraic structure of quantum operators, without any reference to particular quantum states.
There has been a series of debate, however, on whether contextuality identified by BKS theorems has any physical consequence~
\cite{CGA1998,Meyer1999,Kent1999,CK2000,SZWZ2000,HKSS2001,Appleby2002,SBZ2001,Larsson2002,Cabello2002,HLZPG2003}.

On the other hand, state-dependent contextual inequalities prove that there is an experimentally verifiable consequence of contextuality. However state-dependent experiments can not determine whether all quantum states possess contextuality that is experimentally verifiable. This question has been answered in affirmative manner in Ref.~\cite{Cabello} by Cabello.

Suppose $A_{ij}$ are dichotomic observables of a noncontextual hidden variable theory taking the values $\pm1$.  We have the following inequality,
\begin{eqnarray}
&-&\langle A_{12}A_{16}A_{17}A_{18} \rangle -\langle A_{12}A_{23}A_{28}A_{29} \rangle
-\langle A_{23}A_{34}A_{37}A_{39} \rangle  \nonumber \\
&-&\langle A_{34}A_{45}A_{47}A_{48}\rangle -\langle A_{45}A_{56}A_{58}A_{59}\rangle -\langle A_{16}A_{56}A_{67}A_{69} \rangle \nonumber \\
&-&\langle A_{17}A_{37}A_{47}A_{67} \rangle -\langle A_{18}A_{28}A_{48}A_{58} \rangle
-\langle A_{29}A_{39}A_{59}A_{69} \rangle \leq 7.
\label{SIC}
\end{eqnarray}
because, assuming the possibility of outcome assignment, the product of all terms on the left hand side is $1$, regardless of how we assign values to each observable. The labeling of observables $A_{ij}$ will be explained in details in Sec. \ref{sec:statedepandindep}.

On the other hand, let $A_{ij} = 2 |v_{ij}\rangle\langle v_{ij}|-\mathbb{I}$ be a quantum operator on the four-dimensional Hilbert space, where the set of four vectors $|v_{ij}\rangle$ corresponding to observables in every correlation term, $\langle A_{ij}A_{kn}A_{ml}A_{pq}\rangle$,  are mutually orthogonal and are defined in Figure \ref{fig:cabellorefpic} (Sec. \ref{sec:statedepandindep}).  It is easy to see that for these operators each term is equal to the identity operator. This implies the left-hand side of Equation (\ref{SIC}) is equal to $9$ regardless of which quantum state the expectation values are evaluated on.  Therefore, all quantum states of dimension $4$, including the completely mixed state, violate Inequality (\ref{SIC}). This requisite state-independent inequality  proves that the state-independent contextuality discovered by BKS is actually experimentally verifiable. We will come back to this inequality in Section \ref{sec:statedepandindep}.

\subsubsection{BKS theorems and state-independent contextual inequalities}
We have seen two different approaches to proving the contextuality of quantum theory.  One is epitomized by the BKS theorem and the other is based on inequalities.  They both utilize the noncontextuality assumption and the algebraic relation of the observables.  Perhaps it is not surprising that the observables used to prove the BKS theorems yield state-independent contextual inequalities.

In fact, the observables in the inequality above are precisely the ones used in Ref.~\cite{Garcia-Alcaine}.
Quantum operators used in other BKS theorems are also known to provide state-independent contextual inequalities.  For example:
\begin{description}
\item[Peres-Merimin squares~\cite{PM3,PM1,Mermin2} (see also Section \ref{sec:BellKochenSpecker}):]
\begin{eqnarray}
& &\langle P_{14}P_{15}P_{16} \rangle + \langle P_{24}P_{25}P_{26} \rangle +
\langle P_{34}P_{35}P_{36} \rangle \nonumber \\ &+& \langle P_{14}P_{24}P_{34} \rangle
 + \langle P_{15}P_{25}P_{35} \rangle - \langle P_{16}P_{26}P_{36} \rangle \leq 4,
\label{MP-square}
\end{eqnarray}
This inequality is violated up to 6 when the $\pm 1$ observables are identified with the following two-qubit operators,
\begin{align}\label{tab:merminperes}
P_{14} = \sigma_z \otimes \mathbb{I} && P_{15} = \mathbb{I} \otimes \sigma_z && P_{16} = \sigma_z\otimes \sigma_z \nonumber \\
P_{24} = \mathbb{I} \otimes \sigma_x && P_{25} = \sigma_x \otimes \mathbb{I} && P_{26} = \sigma_x\otimes \sigma_x \nonumber \\
P_{34} = \sigma_z \otimes \sigma_x  && P_{35} = \sigma_x \otimes \sigma_z && P_{36} = \sigma_y \otimes \sigma_y.
\end{align}

\item[Mermin's GHZ-type proof~\cite{Mermin2,PM3}:]
\begin{multline}
 \langle A_1 B_1 B_2 \prod_{i=3}^n B_i\rangle + \langle A_2 B_1 C_2 \prod_{i=3}^n C_i\rangle + \langle A_3 C_1 B_2 \prod_{i=3}^n C_i\rangle
 + \langle A_4 C_1 C_2 \prod_{i=3}^n B_i\rangle -\langle A_1 A_2 A_3 A_4 \rangle \leq 3.
\end{multline}
This inequality is violated when each observable is set as the following tensor product of $n$ single-qubit observables,
\begin{align}
 A_1 &= \sigma_z \otimes \sigma_z \otimes \sigma_z \otimes \ldots \otimes \sigma_z\\
 A_2 &= \sigma_z \otimes \sigma_x \otimes \sigma_x \otimes \ldots \otimes \sigma_x\\
 A_3 &= \sigma_x \otimes \sigma_z \otimes \sigma_x \otimes \ldots \otimes \sigma_x\\
 A_4 &= \sigma_x \otimes \sigma_x \otimes \sigma_z \otimes \ldots \otimes \sigma_x\\
 B_i &= \sigma_{z,i}\\
 C_i &= \sigma_{x,i}.
\end{align}
Here, $\sigma_{z,i}$ and $\sigma_{x,i}$ denote the Pauli $Z$ and $X$ operator on the $i$-th qubit, respectively.
\end{description}
These examples are also presented in Ref.~\cite{Cabello}.

Despite these examples, it is not known whether any set of observables used in some BKS theorem yields a state-independent inequality.  Moreover, state-independent contextual inequalities can be constructed from a set of observables that do not allow a BKS theorem~\cite{Bengtsson,Yu2}.
There is much to be investigated about relations between the BKS theorems and state-independent contextual inequalities.

\subsubsection{Reducing the number of measurements}
The set of observables  characterizes a contextual inequality.  The size of this set influences the number of measurements that must be performed.  When fewer observables are required to measure the quantum violation of an inequality the experimental implementation requires fewer experimental resources.  For the case when all the observables are rays (i.e., rank-$1$ projectors), the minimum number of observables is upperbounded by $10 + d$.  For $d=3$, a set of 13 rays has been discovered by Yu and Oh~\cite{Yu2}, which has been proven to be optimal in Ref.~\cite{Cabello2}.

 \begin{figure}[ht!]
  \begin{center}
 \includegraphics[width=0.4\columnwidth]{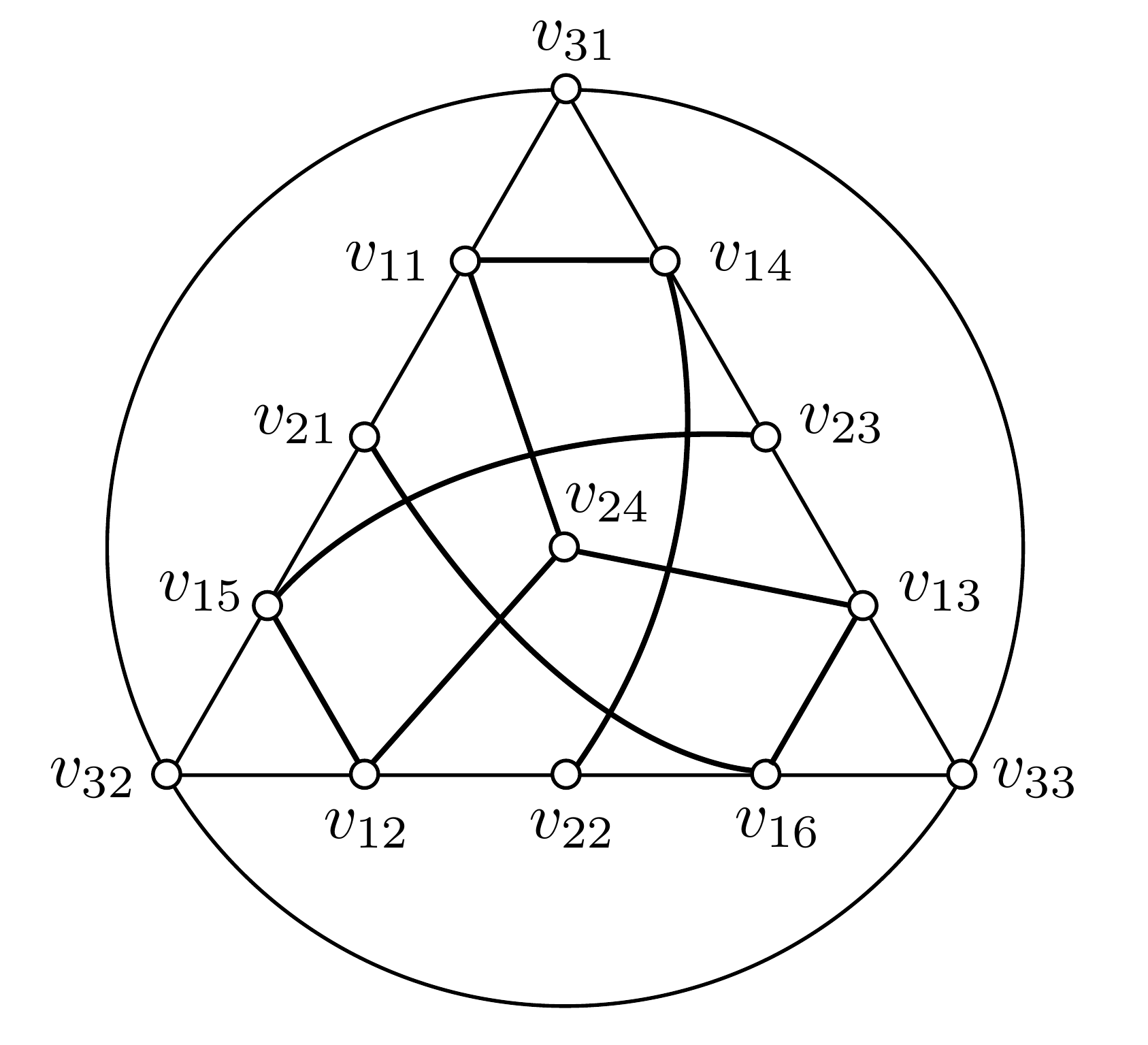}
  \end{center}
  \caption{\small The graph of the complementarity relations for the observables $A_ij$ corresponding to the 13 rays in Equation (\ref{eq:yuandohprojectors}). Vertices represent observables and compatible observables are connected by an edge. Any collection of observables belonging to a clique are mutually compatible. }
  \label{fig:13rays}
\end{figure}

The 13 rays of Yu and Oh form the set $V$ and are of the form
\begin{align}\label{eq:yuandohprojectors}
&v_{11} = (0,1,-1), && v_{21} = (-1,1,1), & v_{31} = (1,0,0), \nonumber \\
&v_{12} = (-1,0,1), && v_{22} = (1,-1,1), & v_{32} = (0,1,0), \nonumber \\
&v_{13} = (1,-1,0), && v_{23} = (1,1,-1), & v_{33} = (0,0,1), \nonumber \\
&v_{14} = (0,1,1),  && v_{24} = (1,1,1),  & ~~ \nonumber \\
&v_{15} = (1,0,1),~~ && ~~ & ~~ \nonumber \\
&v_{16} = (1,1,0).~~ && ~~&
\end{align}
The complementarity relations of the corresponding dichotomous observables $A_{ij}=\mathbb{I} -2|v_{ij}\rangle\langle v_{ij}|$  are indicated in Figure \ref{fig:13rays}.  Any noncontextual assignment of $\pm1$-outcomes to these observables will satisfy the condition:
\begin{equation}\label{eq:13projectorinequality}
\sum_{v_{ij}} \langle A_{ij} \rangle - \frac{1}{4}\sum_{v_{ij},v_{nk}\in V}\Gamma_{(ij)(nk)}\langle A_{ij} A_{nk}\rangle \leq 8,
\end{equation}
where $\Gamma_{(ij)(nk)}=1$ if two rays $v_{ij}$ and $v_{nk}$ are orthogonal; in Figure \ref{fig:13rays} vertices $v_{ij}$ and $v_{nk}$ are adjacent. Otherwise it is equal to zero. The quantum value of the above inequality can reach $25/3$, which is $4 \%$ higher than the noncontextual bound.

To evaluate the quantum violation, some of the contextual observables must be measured sequentially.  The coherence of the quantum system must be maintained from the measurement of the first observable until the last one.  For example, verification of the violation of Equation (\ref{SIC}) requires four observables to be measured sequentially.  The coherence becomes more difficult to maintain as we increase the number of sequential measurements.  Hence, it is desirable to reduce the number of such observables as much as possible.  For dimensions $3,4$ and $5$, it is proven that state-independent contextual inequalities can be constructed with at most two sequentially measured observables~\cite{Yu2,Yu1,Kleinmann}.

\subsubsection{More robust state-independent inequalities against experimental errors}
Suppose we obtain the value $v$ in some experiment when confirming contextuality using a particular inequality.  This could be greater than the classical bound $\aleph$ of the given inequality, but as long as the number of times we repeat an experiment is finite (which is always the case), it is possible that we observed this violation by accident.
If the separation between the classical value and the quantum value is greater, then there is less chance of ``guessing wrong".

Given a set $\mathbf{S}$ of observables, let us denote the family of sets of indices corresponding to all compatible combinations of observables by $\mathbf{C}$ whose elements are $\mathbf{c}$, i.e., $\mathbf{c} \in \mathbf{C}$.
In Equation (\ref{SIC}), $\mathbf{S}$ consists of all 18 variables $A_{ij}$; an example of an element $\mathbf{c}$ is $\{12,16,17,18\}$.

Next, let us denote by $A_{\mathbf{c}}$ a product of compatible $\pm 1$ quantum observables $A_i$, i.e., $A_{\mathbf{c}}=A_i A_j \dots A_k$, where $(i,j,\dots,k)=\mathbf{c}\in\mathbf{C}$. Moreover, let $a_{\mathbf{c}}$ be an analogical product of possible outcome assignments $a_i=\pm 1$, $a_{\mathbf{c}}=a_i a_j \dots a_k$. Finally, let $\lambda_{\mathbf{c}}$ denote some real coefficient.

All the known state-independent contextual inequalities are given as expectation values of a linear combination of $A_{\mathbf{c}}$ with respect to some quantum state $\rho$. The classical bound $\aleph$ of the inequality is determined by a noncontextual model which is based on outcome assignments $a_{\mathbf{c}}$
\begin{equation}\label{eq:Kleinmanninequality}
 \sum_{\mathbf{c} \in \mathbf{C}} \lambda_{\mathbf{c}} a_{\mathbf{c}} \leq \aleph.
\end{equation}
Let $q$ be the quantum value of some given state-independent contextual inequality, i.e.,
\begin{equation}
T(\lambda)_{\rho} = \sum_{\mathbf{c} \in \mathbf{C}} \lambda_{\mathbf{c}} \langle A_{\mathbf{c}} \rangle_{\rho} = q.
\end{equation}
Moreover, let us assume that coefficients $\lambda_{\mathbf{c}}$ are normalized such that $q=1$. In this case, the set of observables $\mathbf{S}$ manifests contextuality if $\aleph<1$. The goal is to find a set of coefficients $\lambda_{\mathbf{c}}$ maximizing the ratio $q/\aleph$ and at the same time obeying $q=1$
\begin{equation}
\max_{\lambda} \frac{T(\lambda)_{\rho}}{\aleph} -1.
\end{equation}

Klienmann and others presented a linear programming method to maximize the ratio $q/\aleph$~\cite{Kleinmann}. Their method assumes that $\mathbf{S}$ and $\mathbf{C}$ are first given.  They set $q=1$ and minimize $\aleph$ by varying $\lambda_\mathbf{c}$ in linear programming.  They argue that the same experimental data obtained in Ref.~\cite{ZWDCLHYD12} can be used to increase the standard deviation of the quantum violation from 3.7 to 7.5, if they use their optimized inequality instead of the unoptimized one presented in Ref.~\cite{Yu2}.

\subsubsection{Implementability under specific quantum systems}

State-independent inequalities can be customized for specific quantum systems.  In quantum optics, the quadratures of the electromagnetic field can be measured with near-unit efficiency using homodyne measurements.  These quadratures are continuous variables, but all state-independent contextual inequalities mentioned above use discrete variables.  Plastino and Cabello have addressed this point and presented the following state-independent contextual inequality with continuous variable~\cite{Plastino}.

State-independent contextual inequality are generally functions of real observables.  On the other hand, the inequality presented in Ref.~\cite{Plastino} is a complex function instead.  Let $x_1$ and $x_2$ be position operators on two different subsystems, and $p_1$ and $p_2$ be the canonically conjugate momentum operators.  These operators satisfy the canonical commutation relation $[x_k,x_l]=[p_k,p_l]=0$ and $[x_k,p_l]=\mathrm{i}\hbar \delta_{kl}$.  Consider the following nine complex functions,
\begin{align}
 A_{11} &= A'_{11} + \mathrm{i} A^{''}_{11} & A_{12} &= A'_{12} + \mathrm{i} A^{''}_{12} & A_{13} &= A'_{13} + \mathrm{i} A^{''}_{13}\\
 A_{21} &= A'_{21} + \mathrm{i} A^{''}_{21} & A_{22} &= A'_{22} + \mathrm{i} A^{''}_{22} & A_{23} &= A'_{23} + \mathrm{i} A^{''}_{23}\\
 A_{31} &= A'_{31} + \mathrm{i} A^{''}_{31} & A_{32} &= A'_{32} + \mathrm{i} A^{''}_{32} & A_{33} &= A'_{33} + \mathrm{i} A^{''}_{33},
\end{align}
where $\mathrm{i}$ is the imaginary number, and $A'_{kl}$ and $A^{''}_{kl}$ are real-valued observables defined by
\begin{align}
 A'_{11} &= \cos \Bigl(\frac{p_0}{\hbar} x_1 \Bigr) & A^{''}_{11} &= \sin \Bigl(\frac{p_0}{\hbar} x_1 \Bigr) \\
 A'_{12} &= \cos \Bigl(\frac{\pi}{p_0} p_2 \Bigr) & A^{''}_{12} &= \sin \Bigl(\frac{\pi}{p_0} p_2 \Bigr) \\
 A'_{13} &= \cos \Bigl(\frac{p_0}{\hbar} x_1 + \frac{\pi}{p_0} p_2 \Bigr) & A^{''}_{13} &= \sin \Bigl(\frac{p_0}{\hbar} x_1 + \frac{\pi}{p_0} p_2 \Bigr) \\
 A'_{21} &= \cos \Bigl(\frac{p_0}{\hbar} x_2 \Bigr) & A^{''}_{21} &= -\sin \Bigl(\frac{p_0}{\hbar} x_2 \Bigr) \\
 A'_{22} &= \cos \Bigl(\frac{\pi}{p_0} p_1 \Bigr) & A^{''}_{22} &= \sin \Bigl(\frac{\pi}{p_0} p_1 \Bigr) \\
 A'_{23} &= \cos \Bigl(\frac{p_0}{\hbar} x_2 - \frac{\pi}{p_0} p_1 \Bigr) & A^{''}_{23} &= \sin \Bigl(\frac{p_0}{\hbar} x_2 - \frac{\pi}{p_0} p_1 \Bigr) \\
 A'_{31} &= \cos \Bigl(\frac{p_0}{\hbar} (x_2 - x_1) \Bigr) & A^{''}_{31} &= \sin \Bigl(\frac{p_0}{\hbar} (x_2 - x_1) \Bigr) \\
 A'_{32} &= \cos \Bigl(\frac{\pi}{p_0} (p_1 + p_2) \Bigr) & A^{''}_{12} &= -\sin \Bigl(\frac{\pi}{p_0} (p_1 + p_2) \Bigr) \\
 A'_{33} &= \cos \Bigl(\frac{p_0}{\hbar} (x_1 - x_2) + \frac{\pi}{p_0} (p_1 + p_2) \Bigr) & A^{''}_{33} &= \sin \Bigl(\frac{p_0}{\hbar} (x_1 - x_2) + \frac{\pi}{p_0} (p_1 + p_2) \Bigr).
\end{align}
In noncontextual hidden variable theories, the variables $A_{kl}$ are modular variables, which implies that
\begin{multline}
 |\langle A_{11} A_{12} A_{13} \rangle + \langle A_{21} A_{22} A_{23} \rangle + \langle A_{31} A_{32} A_{33} \rangle + \langle A_{11} A_{21} A_{31} \rangle\\
 + \langle A_{12} A_{22} A_{32} \rangle - \langle A_{13} A_{23} A_{33} \rangle | \leq 3 \sqrt{3}.
\end{multline}
In quantum theory, we obtain the value 6 for the left hand side, therefore we have a state-independent contextuality for continuous variables.

\subsection{State-independent to state-dependent}\label{sec:statedepandindep}

Let us now discuss the relation between state-dependant and state-independent contextuality tests. An interesting analysis of the transition between state-independent and state-dependent inequalities was done in Ref. \cite{Cabello}, in following this work we come back to Inequality (\ref{SIC}) for 18 binary $\pm 1$ observables $A_{ij}$ on a four-level system. We present the analysis below. Each observable appears in two different measurement scenarios (contexts) that are labeled by indices $i,j=1,\dots,9$ that obey $i<j$. In total there are nine different contexts. Two observables $A_{ij}$ and $A_{kl}$ can be measured together if they share the same context. For convenience we rewrite the state-independent inequality (\ref{SIC}):
\begin{eqnarray}
&-&\langle A_{12}A_{16}A_{17}A_{18} \rangle -\langle A_{12}A_{23}A_{28}A_{29} \rangle
-\langle A_{23}A_{34}A_{37}A_{39} \rangle  \nonumber \\
&-&\langle A_{34}A_{45}A_{47}A_{48}\rangle -\langle A_{45}A_{56}A_{58}A_{59}\rangle -\langle A_{16}A_{56}A_{67}A_{69} \rangle \nonumber \\
&-&\langle A_{17}A_{37}A_{47}A_{67} \rangle -\langle A_{18}A_{28}A_{48}A_{58} \rangle
-\langle A_{29}A_{39}A_{59}A_{69} \rangle \leq 7. \nonumber
\end{eqnarray}

 \begin{figure}[ht!]
  \begin{center}
 \includegraphics[width=0.5\columnwidth]{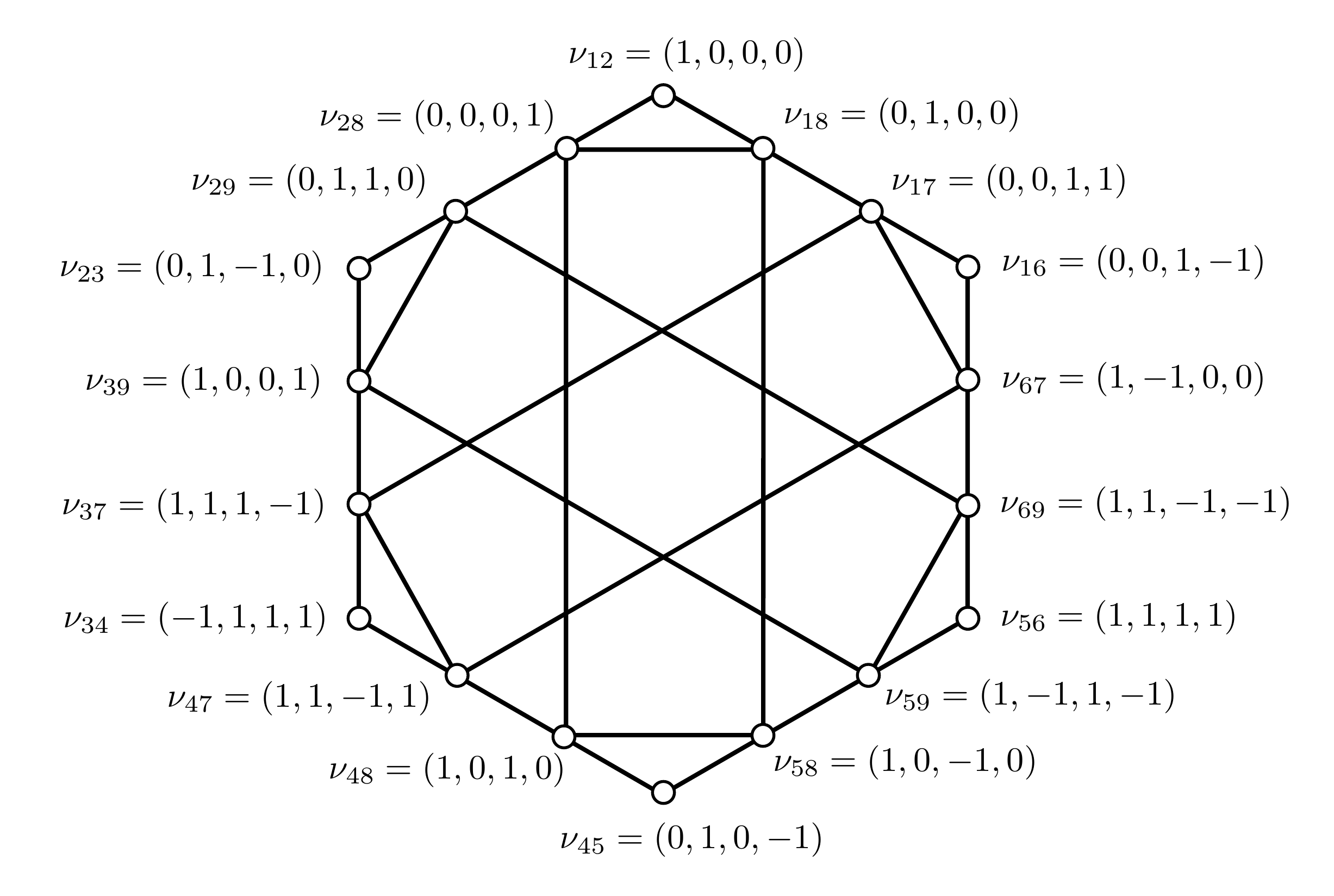}
  \end{center}
  \caption{\small The graph $G$ of the compatibility relations of the observables in Inequality (\ref{SIC}).}
  \label{fig:cabellorefpic}
\end{figure}

Next, if one assigns $\pm 1$ values to 13 observables $A_{ij}$ in Equation (\ref{SIC}) one can arrive at the inequality
\begin{equation}
-\langle A_{12}A_{18}\rangle-\langle A_{12}A_{23}\rangle-\langle A_{23}A_{34}\rangle
-\langle A_{34}A_{48}\rangle-\langle A_{18}A_{48}\rangle \leq 3
\end{equation}
that is exactly of the KCBS form \cite{Klyachko}. This inequality is state-dependent. The transition between state-independence and state-dependence occurred because a number of measurements was reduced.

Moreover, let us recall the state-independent inequality (\ref{MP-square}) that was introduced in Ref. \cite{Cabello}
\begin{eqnarray}\label{eq:cabelloexperimentcorespondance}
& &\langle P_{14}P_{15}P_{16} \rangle + \langle P_{24}P_{25}P_{26} \rangle +
\langle P_{34}P_{35}P_{36} \rangle \nonumber \\ &+& \langle P_{14}P_{24}P_{34} \rangle
 + \langle P_{15}P_{25}P_{35} \rangle - \langle P_{16}P_{26}P_{36} \rangle \leq 4,\nonumber
\end{eqnarray}
where $P_{ij}$ are $\pm 1$ two-qubit observables that are compatible if the indices of $P_{ij}$ and  $P_{kl}$ are related by $i = k$ or $j = l$. This inequality is based on the contextuality proof due to Peres and Mermin \cite{PM3,PM1,Mermin2}. There are nine observables, but if one assigns $\pm 1$ to five of them one can get
\begin{equation}
\langle P_{14}P_{16} \rangle + \langle P_{24}P_{26} \rangle + \langle P_{14}P_{24} \rangle - \langle P_{16}P_{26} \rangle \leq 2.
\end{equation}
This is exactly the CHSH inequality \cite{Clauser} which is state-dependent due to the fact that it contains fewer measurements than the original inequality.

\section{Different approaches to contextuality}
\label{sec:differentapproaches}

We have identified contextuality and nonlocality with the lack of a joint probability distribution capable of reproducing the measurable probabilities as marginals. Using this approach some nonlocal features can be naturally mapped to contextuality. Now that we have developed a broader perspective for the field of contextuality, including a grasp of contextual inequalities, we explicitly discuss several instances of this interdisciplinary approach.

\subsection{Entropic tests}
An alternative approach to testing contextuality is based upon information theoretic principles \cite{Ramanathan, Braunstein, Cerf, Fritz}. This approach was inspired by an information theoretic generalization of the CHSH inequality \cite{Braunstein}.  It uses the existence of a joint probability distribution (governing measurement outcomes) for any systems with non-contextual local hidden variables, to derive relations on the information carried by the system. Here we will describe the direct application off this technique to the KCBS inequality \cite{Ramanathan}.

The backbone of this work is the existence of a joint probability distribution (in non-contextual hidden variable theories) describing the outcomes of five observables $A_1, \dots, A_5$. This facilitates the construction of a joint entropy $H(A_1, A_2, \dots, A_5)$ where
\begin{equation}
H(A) = -\sum_a P_A( a){\rm log}\left[P_A( a)\right],
\end{equation}
is the Shannon entropy. This quantity has several important properties.

There is a simple relation for the amount of information, $H(A|B)$, needed to describe the outcomes of an observable $A$ conditioned on already knowing the value of a second observable $B$:
\begin{equation}
H(AB) = H(A|B) + H(B).
\end{equation}
The second property
\begin{equation}
H(A|B)\le H(A)\le H(AB),
\label{eq:shanonentropyrule2}
\end{equation}
has an intuitive interpretation as conditioning can not increase the information carried by an observable $A$ and that a single observable can never carry more information than a pair of observables.

Using the first of these properties we can determine that

\begin{align}
H(A_1, \dots, A_5) \, = \,H(A_1| A_2,\dots, A_5) \, + \, H(A_2| A_3, A_4, A_5) \, + \\  H(A_3| A_4, A_5) \, + \, H(A_4| A_5)\, + \, H(A_5)
\end{align}

Further coarse-graining this relation using Equation (\ref{eq:shanonentropyrule2}) gives the information theoretic inequality
\begin{equation}
H(A_1|A_5) = H(A_1| A_2) + H(A_2| A_3) + H(A_3| A_4) + H(A_4| A_5)
\end{equation}
This inequality is maximally violated by a value of $0.091$ bits when the state $|\psi\rangle$ and projectors $|A_1\rangle, \dots, |A_5\rangle$ have the relative orientation
\begin{eqnarray}
&&|\psi\rangle  =  \left(\begin{matrix} \sin\theta\\  \cos\theta\\ 0\end{matrix}\right)\qquad |A_1\rangle = \frac{1}{\sqrt{2}}\left(\begin{matrix}\frac{\sqrt{\cos 2\varphi}}{\cos\varphi}\\\tan\varphi\\ 1\end{matrix}\right) \nonumber \\
&&|A_2\rangle = \left(\begin{matrix} 0\\ \cos\varphi\\ -\sin\varphi\end{matrix}\right) \qquad |A_3\rangle = \left(\begin{matrix} 1\\ 0\\ 0\end{matrix}\right), \nonumber \\
&&|A_4\rangle = \left(\begin{matrix}  0\\  \cos\varphi\\ \sin\varphi\end{matrix}\right)\qquad
|A_5\rangle = |A_1\rangle \times |A_4\rangle, \nonumber
\end{eqnarray}
for $\varphi = 0.1698$ and $\theta = 0.2366$. This implies that no joint probability distribution exists for the outcomes of the five observables on the state $|\psi\rangle$.

\subsection{Monogamy relation}

As  a counterpoint to our discussion of deriving contextual inequalities via the graph independence number of the corresponding complementarity graph; we look at the roles of other graph properties mentioned in Subsection \ref{sec:contextualrelationsbounds}. Specifically, we mentioned that in principle it is possible to violate contextuality by more than it is allowed by the Lovasz $\vartheta$-function (quantum limit). The maximal contextuality compliant with the no-disturbance principle is given by the fractional packing number of a complementarity graph of the corresponding inequality \cite{CSW}. In nonlocal scenarios no-disturbance is equivalent to no-signaling. Here we consider an approach to contextuality based upon this relationship.

The no-signaling condition was used in Ref. \cite{Monogamy1} to derive bounds on violations of Bell inequalities in scenarios in which one observer takes part in violations of more than one inequality of the same type. It was found that violation of corresponding inequalities depends on the number of observers and the number of measurement settings. For example, in every no-signaling theory in the CHSH scenario in which Alice, Bob and Charlie can choose one of two measurements each, the sum of violations of CHSH inequalities between Alice and Bob and between Alice and Charlie is bounded by four
\begin{equation}
\langle CHSH_{AB}\rangle+\langle CHSH_{AC}\rangle \leq 4.
\end{equation}
Therefore, if the inequality is violated between Alice and Bob it cannot be violated between Alice and Charlie, and vice versa. This property is refered to as the monogamy of Bell inequality violations. If both inequalities were violated, the cooperation between Bob and Charlie (who do not have to be spatially separated) could lead to a superluminal signaling to Alice.

In Ref. \cite{Monogamy2} monogamy relations were derived for KCBS inequalities. This time the monogamy is based on no-disturbance. The KCBS scenario involves five cyclically exclusive events $A_1,\dots, A_5$. Next, imagine another KCBS scenario involving measurements $A'_1,\dots,A'_5$. However, assume that measurements in the first set are not independent form the measurements in the second set. This is in a sense analogical to the CHSH monogamy scenario in which two CHSH inequalities are not independent, because they are related via Alice's measurements. There are different ways to implement the dependence between events $A_i$ and $A'_j$. For example, some of these measurements may be common to both sets, or they might be related by additional exclusivity relations.

\begin{figure}[hbt!]
\begin{center}
\includegraphics[width=0.7\columnwidth]{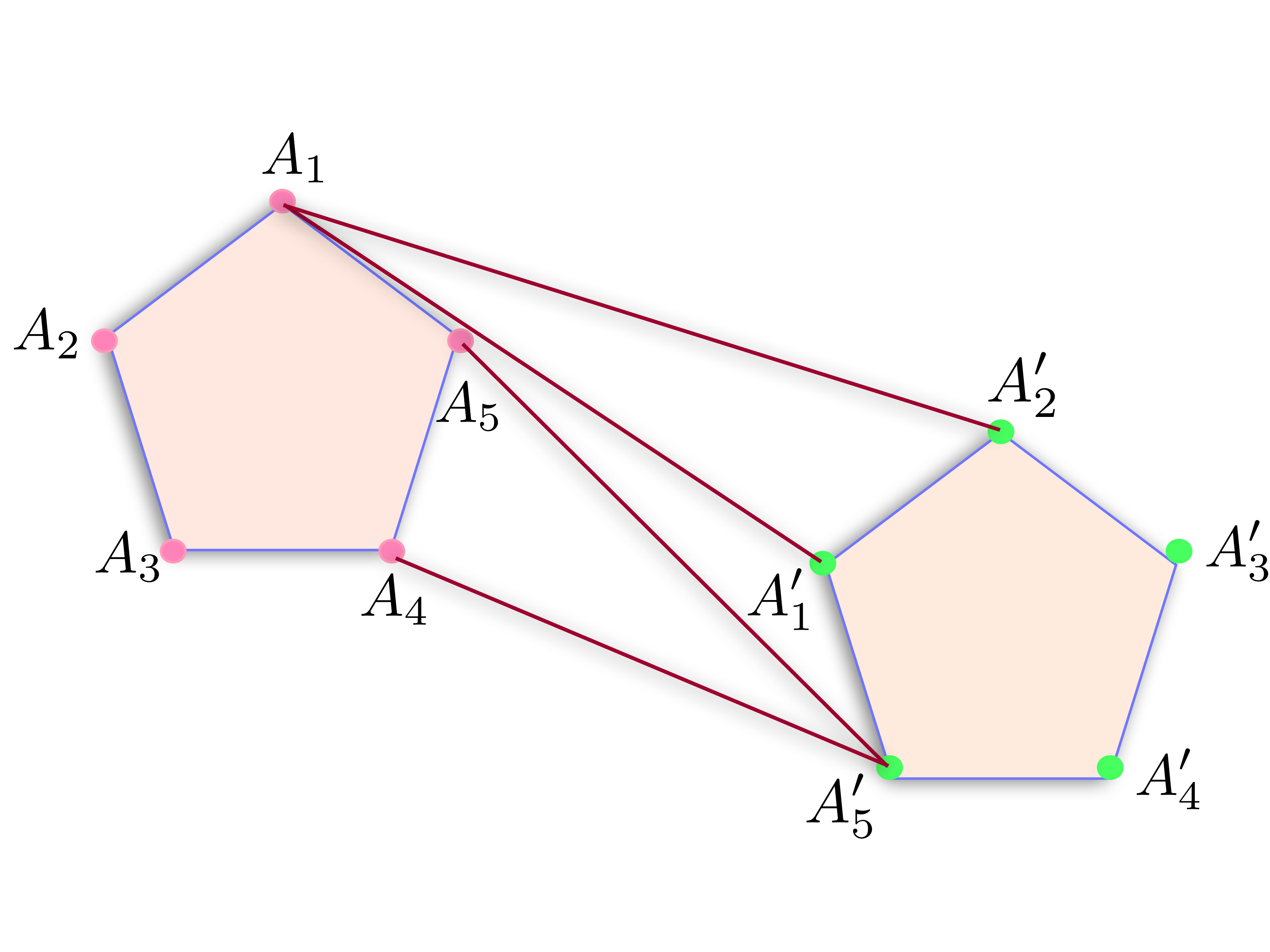}
\caption{\small An example of exclusivity graph leading to the monogamy of violations of two KCBS inequalities.}
\label{fig:monogamy}
\end{center}
\end{figure}

In Figure \ref{fig:monogamy} we present one possibility in which event $A_1$ is exclusive to $A'_1$ and $A'_2$, and event $A'_5$ is exclusive to $A_4$ and $A_5$.
Therefore, in addition to $p(A_1=1,A_5=0)$ and $p(A_1=1,A_2=0)$ one can also experimentally estimate $p(A_1=1,A'_1=0,A'_2=0)$ (similar for $A'_5$). We set $p(A_1=1)=p$ and $p(A'_5=1)=q$. Exclusivity implies that $p(A'_1=1)+p(A'_2=1)\leq 1-p$ and $p(A_4=1)+p(A_5=1)\leq 1-q$. Moreover, $p(A'_3=1)+p(A'_4=1)\leq 1$ and $p(A_2=1)+p(A_3=1)\leq 1$. As a result
\begin{equation}
\sum_{i=1}^5 p(A_i=1) \leq 2+p-q,~~\sum_{j=1}^5 p(A'_j=1) \leq 2-p+q,
\end{equation}
and
\begin{equation}
\sum_{i=1}^5 p(A_i=1) + \sum_{j=1}^5 p(A'_j=1) \leq 4.
\end{equation}
Due to additional exclusivity relations only one of the two KCBS inequalities can be violated. This effect can be interpreted as a monogamy of contextuality. If no-disturbance was not obeyed and $p(A_1=1,A_5=0)$ or $p(A_1=1,A_2=0)$ were different from $p(A_1=1,A'_1=0,A'_2=0)$ (or similar for $A'_5$), then monogamy relation would not have to hold.

\section{Experiment}\label{sec:experiment}
Experimental tests of quantum contextuality  have developed rapidly over the last few years. In general these experiments aim to physically realize violation of a contextual inequality within a quantum system. Some examples of these inequalities are given in Section \ref{sec:contextualityinequalities}

\begin{figure}
\centerline{\scalebox{0.45}{\includegraphics{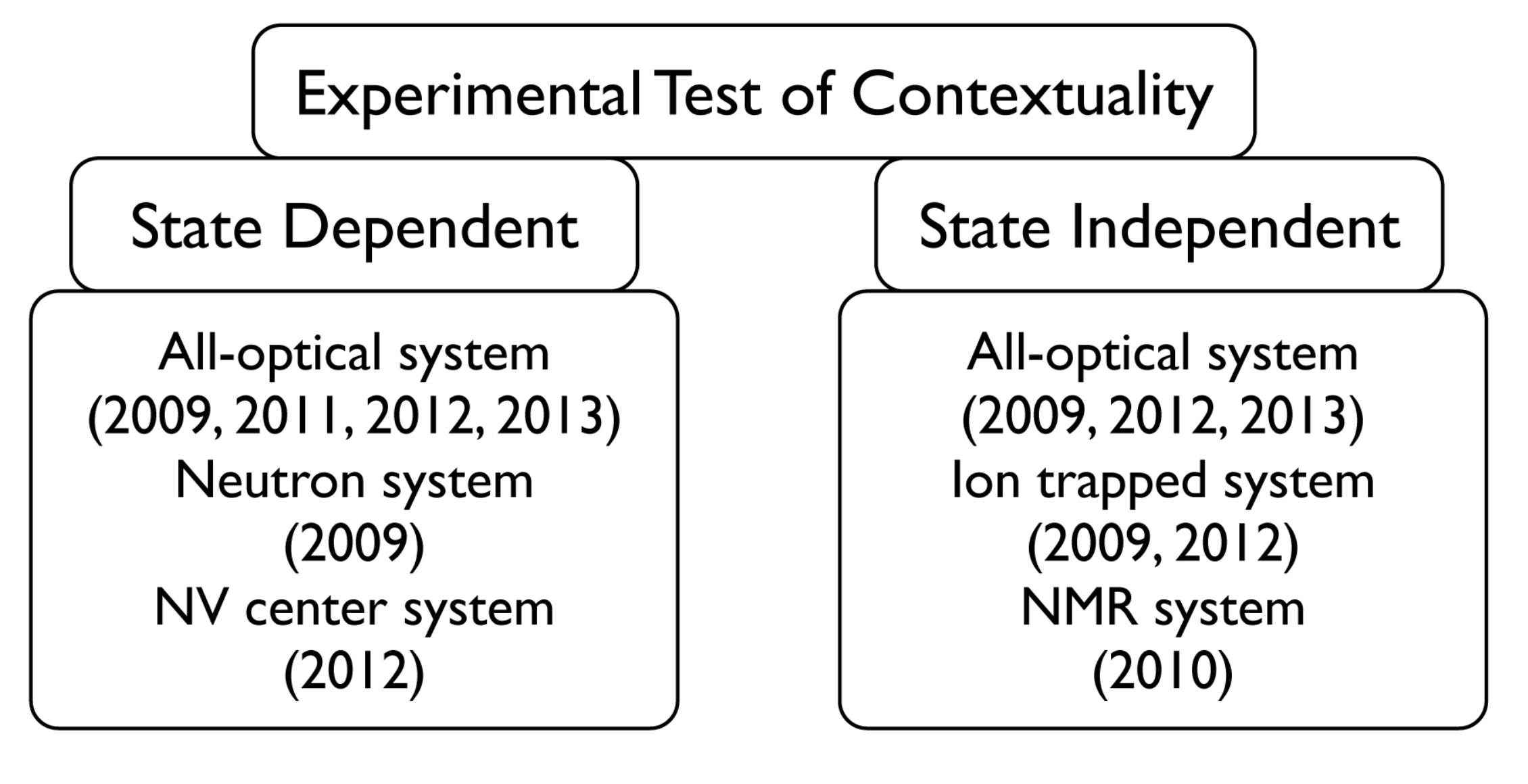}}}
\caption{An Outline of experimental tests of quantum contextuality circa 2009. }
\label{fig:fig1}
\end{figure}

Experimental realizations of contextual inequalities have appeared in a diverse array of physical systems, including but not limited to:  all-optical experiments, neutrons,  nitrogen-vacancy (NV) centers, ion traps and nuclear magnetic resonance (NMR) systems. To capture the rapid and dynamical development of this field we emphasis experiments from 2009 and try to give a feeling for how progressive experiments have extended and consolidated results. For a concise summary of the many different parallel avenues of development see Figure \ref{fig:fig1}.

\subsection{Tests of the Mermin-Peres square proof of quantum contextuality}

\subsubsection{State-independent realizations}
Early experimental work was conducted on variants of Inequality (\ref{MP-square}). This state-independent inequality for two qubit systems, is a compact way of verifying the lack of preassigned  noncontextual measurement outcomes for the Mermin-Peres  square operators in Equation (\ref{tab:merminperes}) \cite{Peres, Mermin1, Mermin2}. The inequality is state-independent because for every triad of operators belonging to a row or column of the Mermin-Peres square it is impossible to preassign noncontextual $\pm1$-outcomes in such a way that the product of operators is equal to the product of the assigned outcomes.

Originally Kirchmair et al. \cite{KZGKGCBR09} focussed on testing this inequality with a pair of  $^{40}$Ca$^{+}$ ions in a linear Pauli trap. Two ionic energy levels furnished a two level systems. These were defined as the basis states $|\uparrow\rangle $ and $|\downarrow\rangle$ of the Pauli operator $\sigma_z$ in Equation (\ref{MP-square}).

Initially the system was prepared in the singlet state:
\begin{equation}
|\psi\rangle = \frac{1}{\sqrt{2}} \left(|\uparrow \downarrow\rangle - |\downarrow \uparrow\rangle \right).
\end{equation}
Constructively measuring the triad of operators belonging to a row or column of the Mermin-Peres square on a collection of $6,600$ copies of $|\psi\rangle$, yielded an average violation of $5.46$.

Further experimental runs were preformed for nine other quantum states of different purity and entanglement. The measured violation for each of these states was between $5.23$ and $5.46$, regardless of the temporal order in which the triads of operators (belonging to a single row or single column of the Mermin-Peres square) were sequentially measured.

The high fidelity of state detection made this experiment decisive in closing detection loopholes. However the observables were not perfectly compatible. This means noncontextual hidden variables (NCHVs) could have been disturbed during the sequential measurement process. Thereby allowing a NCHV model to have context dependent measurement statistics and making violation of the contextual inequality insufficient evidence to exclude a joint probability distribution. To address this issue Kirchmair et al. tested a version of the inequality which could decisively exclude disturbed hidden variables.

Concurrent experimental work on variants of Inequality (\ref{MP-square}) was implemented in all optical and nuclear magnetic resonance systems. All optical experiments offer a platform where the no-disturbance condition can be reliably satisfied, however the detection loophole is a more prominent issue. Here the detection loophole requires us to make auxiliary assumptions about the putative measurement outcomes of lost photons. We must assume that the  measurement statistics collected during successful runs fairly representant both detected and lost photons. This becomes an implicit assumption when a statistically significant proportion of photons are lost.

An interesting development in all optical implementations is that all four levels of the two qubit system can be encoded in a single photons. This requires both path and polarization modes of the photon to be used. Both degrees of freedom span a $2$-dimensional Hilbert space. One realization would be to introduce a beam splitter and use the encoding
\begin{align}\label{eq:encodinghilbertspace}
&|0\rangle=|t,H\rangle, &|1\rangle=|t,V\rangle, \qquad & |2\rangle=|r,H\rangle &&|3\rangle=|r,V\rangle,
\end{align}
 where $t$ and $r$ are the transmitted and reflected paths of the photon, while $H$ and $V$ are two independent linear polarization states. This development makes it possible to test contextuality of an indivisible quantum system, thereby guaranteeing that any violation of Inequality (\ref{MP-square}) can not be due to entanglement between the two qubits. An all optical experimental test of this inequality was carried out by Anselmen et al. For a collection of  20 different single-photon states, they observed  violation of the inequality by  $5.4521$ on average \cite{ARBC09}.

Corroboration of these results was found by O. Moussa et al. who used nuclear magnetic resonance techniques to test the inequality for an ensemble of nuclear spins in a macroscopic single crystal of malonic acid ($C_3H_4O_4$) \cite{MRCL10}.
A small fraction ($3\%$) of the molecules in this solid state sample were triply labeled with $^{13}C$ to form an ensemble of processor molecules. In this environment violation up to  $5.3$ was detected.

\subsubsection{State-dependent realizations.}

In Subsection \ref{sec:statedepandindep} we explained how to go from state-independent to state-dependent inequalities by restricting the number of observables. These results have been applied to experimental work on Mermin-Peres square operators. Here we highlight an experiment based on a state-dependent inequality derived from Equation  (\ref{MP-square}) by a priori assigning values of $\pm 1 $ to certain observables.  The theoretical process described in Subsection \ref{sec:statedepandindep} is physically implemented by restricting the range of measurement settings.

To illustrate these principles we consider an all optical experiment based on the inequality:
\begin{eqnarray}\label{eq:modifiedmerminperes}
\langle P_{14} P_{15}P_{16}\rangle+\langle P_{24}P_{25}P_{26}\rangle+\langle P_{34} P_{35}\rangle+\langle P_{14} P_{24} P_{34}\rangle \nonumber\\
+\langle P_{15}P_{25}P_{35}\rangle- \langle P_{16}P_{26}\rangle \leq 4,
\end{eqnarray}
which can be  derived directly from Equation (\ref{MP-square}) by preassigning the value of $P_{36} = +1$  \cite{LHGSZLG09}. This inequality was experimentally tested on a two qubit system constructed from the polarization degrees of freedom of a pair of entangled photons \cite{LHGSZLG09}. The state preparation phase of the experiment used spontaneous parametric down-conversion to generate two-photon product states  which were subsequently manipulated by standard optical elements into the two-qubit state:
\begin{equation}
|\psi\rangle = \left[\cos{\frac{3\pi}{8}}|H\rangle+\sin{\frac{3\pi}{8}}|V\rangle\right]\otimes \left[\cos\frac{15\pi}{8}|H\rangle+\sin\frac{15\pi}{8}|V\rangle\right],
\end{equation}
where $|H\rangle $ and $|V\rangle$ are horizontal and vertical polarization states. While the measurement process was implemented by the parity check technique, utilizing polarization beam splitters and half wave plates. The state $|\psi\rangle$ saturates the quantum bound of Equation (\ref{eq:modifiedmerminperes}) placing a theoretical upper bound of $5$ on the detectable violation. In practice during the experiment, the Hong-Ou-Mandel interference visibility was about $85\%$ such that the maximum violation achieved was $4.85$.

Similar principles were experimentally tested on a single neutron system entangled in spin and path \cite{BKSSCRH09}. The experiment focused on a collection of six observables given by Pauli operators on the $2$-dimensional Hilbert spaces spanned by the neutron spin (s) and path (p) degree of freedom respectively. The corresponding contextual inequality for these $6$ operators is:
\begin{eqnarray}
-\langle \sigma^s_x\cdot \sigma^p_x\rangle-\langle \sigma^s_y\cdot \sigma^p_y\rangle-\langle \sigma^s_x \sigma^p_y\cdot \sigma^s_y \sigma^p_x \rangle\leq 1.
\end{eqnarray}
A neutron beam with a mean wavelength $1.92 \text{\AA}$ was selected by a Si perfect crystal monochromator,
 to create a source of single neutrons which were subsequently prepared in maximally entangled Bell-like state (in the path and spin degrees of freedom). In the single neutron system the maximum experimental violation of the inequality was $2.299$.

These experimental implementations form an extensive body of evidence supporting the statement: it is not possible to preassigned noncontextual measurement outcomes to the Mermin-Peres square observables.

\subsection{A minimal state-independent test of contextuality}

Contiguous experimental tests of quantum contextuality have been engineered from other contextual inequalities. We briefly highlight these endeavors by following a sequence of state-independent experiments. We will subsequently tackle their state-dependent counterparts.

To be theoretically capable of preforming a state-independent test of quantum contextuality on a $d$-dimensional system, you need at least $(10 + d)$ rank-$1$ projectors \cite{Cabello13}. It follows that the minimal state-independent test of quantum contextuality requires $13$ observables since $3$-dimensional systems are the smallest contextual systems. A condition satisfied by Inequality (\ref{eq:13projectorinequality}). This has inspired a body of theoretical and experimental work on  Yu and Oh's 13 projector inequality \cite{Yu2}.

In the experimental arena  C. Zu et al. \cite{ZWDCLHYD12} tested the inequality on a single photonic qutrit. The three level system was constituted by three different spatial paths of a photon.  A single  800nm photon was conditionally obtained by a spontaneous parametric down conversion process and subsequently manipulated using half-wave plates and polarization beam splitters. Theoretically Inequality (\ref{eq:13projectorinequality}) is violated by all qutrits states which uniformly attain the quantum bound of $25/3$. This quantum bound is only $4\%$ beyond the classical bound imposed by non-contextual hidden variable theory; attesting to the high precision attained by C. Zu et al. who successfully demonstrated experimental violation for a collection of quantum states \cite{ZWDCLHYD12}.

In the theoretical arena the 13 projector state-independent inequality was further developed by Kleinmann et al. \cite{Kleinmann}.  Here the emphasis was on developing a tight variant of Inequality (\ref{eq:13projectorinequality}). While violation of a contextual inequality decisively refutes the existence of noncontextual hidden variables; compliance with a full complement of tight inequalities furnish sufficient conditions to prove noncontextual hidden variables exists. Kleinmann et al. developed two tight variants of this inequality which have the general format of Equation (\ref{eq:Kleinmanninequality}). One advantage of the tight inequality is the ten percent gap between the noncontextual hidden variable bound of $25$ and the quantum bound of $25 +8/3$; making it easier to experimentally demonstrate violation of this inequality.

Such an experimental implementation was undertaken using a single $^{171}$Yb$^+$ ion trapped in a four-rods radio frequency trap \cite{ZUZAWDSDK12}. The three internal atomic energy levels of ground state of the ion are mapped directly to the three levels of the qutrit system. For eleven different states, X. Zhang et al. showed violation of the inequality by as much as $27.38$ \cite{ZUZAWDSDK12}.
Again the results from this ion trap experiment were free from the detection loophole but could not directly close the disturbed hidden variables loophole.

To finish our discussion of state-independent experiments we hight an interesting parallel development by  V. D'Ambrosio et al. \cite{DHANBSC13} who experimentally tested the state-independent inequality:
\begin{eqnarray}\label{eq:cabellosexperiment}
\sum_{v_{ij} \in V(G)}P(|v_{ij}\rangle\langle v_{ij}| =1)\leq 4,
\end{eqnarray}
where $V(G)$ is the vertex set  of Figure \ref{fig:cabellorefpic}. Consider a collection of eighteen events, $|v_{ij}\rangle \langle v_{ij}| = 1$, with exclusivity relations summarized by (an exclusivity graph isomorphic to) Figure \ref{fig:cabellorefpic}. Equation (\ref{eq:cabellosexperiment}) is tantamount to the statement that in any noncontextual hidden variable theory at most four of these events can happen at the same time. We note that, this inequality can be derived by rewriting Equation (\ref{SIC}) in terms of the projectors $2|v_{ij}\rangle \langle v_{ij}| = 1 + A_{ij}$. Furthermore it has the fascinating interpretation of being the analogue of Wright's inequality (\ref{eq:wright}) for this state-independent scenario.

V. D'Ambrosio et al. \cite{DHANBSC13} experimentally violated this inequality by $4.517$ on average for 15 states of a four level quantum system encoded in the path and polarization degrees of freedom of a single photon; an amount slightly above the quantum bound for Equation (\ref{eq:cabellosexperiment}). This discrepancy arose because the exclusivity relations were not perfectly implemented in the experiment, causing a small level of coincidence of exclusive events.

\subsection{Minimal tests of contextuality: the KCBS inequality}

It remains to discuss state-dependent experimental tests of quantum contextuality. This scenario is especially interesting because it covers experimental tests with the fewest measurement settings.

The minimum number of observables capable of exhibiting contextuality is five. This follows from the analysis of complementarity relation graphs in Subsection \ref{sec:complementarity} which  demonstrated that $4$ and $5$-cycle complementarity graphs describe the simplest and smallest collections of contextual observables. In the four observable scenario -  where the complementarity relations induce the CHSH inequality for a pair of qubits -  violation of the contextual inequality is usually attributed to entanglement. Therefore in a single quantum system the pursuit of testing quantum contextuality using the least measurement settings has focused on the five observable KCBS inequality (\ref{eq:klyachkoccorrelationineq}).

The first experimental tests of the KCBS inequality used a single photon heralded source and standard optical elements to create a three level system encoded in a subspace of Equation (\ref{eq:encodinghilbertspace}) \cite{LLSLRWZ11}. The  KCBS inequality requires all five observables to be measured in two different contexts, this was implemented using a triplet of detectors which simultaneously measured pairs of observables belonging to the correlation terms $\langle A_i A_j\rangle$ in Equation (\ref{eq:klyachkoccorrelationineq}). However for a specific observable, $A_1$, there was disruption to the state of the observable between the points at which the two different contextual measurements occurred. This forced R. Lapkiewicz et al. to test a variant of the KCBS inequality
\begin{eqnarray}
\langle A_1A_2\rangle+\langle A_2A_3\rangle+\langle A_3A_4\rangle+\langle A_4A_5\rangle+\langle A_5A_1^{'}\rangle \geq -3-\epsilon,
\end{eqnarray}
which included an experimentally measurable error term,  $\epsilon=1-\langle A_1^{'}A_1\rangle$, reflecting the discrepancy between the state of the observable $A_1$ measured in the context of $A_2$ and the state $A_1'$ measured in the context of $A_5$ \cite{LLSLRWZ11}. In practice this error was $\epsilon=-0.083$. A violation of $-3.899$ was attained.

Recently J. Ahrens et al. experimentally implemented the KCBS inequality without introducing any error terms  \cite{AACB13}. The all optical scheme was based on sequential measurements or pairs of compatible observables, $A_i, A_j$ in Equation (\ref{eq:klyachkoccorrelationineq}). This allowed the compatability of observables to be checked by inverting the order of sequential measurement and checking whether the outcome (level of violation of the KCBS inequality) was effected \cite{AACB13} .
The obtained experimental violation was $-3.902$ which is just below the quantum bound of $4\sqrt{5} - 5$ for the KCBS inequality. Due to the close relationship between the KCBS inequality and Wright's inequality (\ref{eq:wright}) this experiment simultaneously tested Wright's inequality. The experimental violation of Wright's inequality, $2.32$, was slightly higher than the maximum quantum bound, $\sqrt{5}$. Again this circumstance was engendered by the lack of perfect exclusivity in the experimental implementation, resulting in a low level of coincidence between mutually exclusive events.

These results were corroborated by Xi Kong et al. who tested Wright's formulation of the KCBS inequality (\ref{eq:wright}) using a single negatively charged nitrogen-vacancy (NV) center in diamond \cite{KSSWHZJDYD12}.
The NV center was composed of an impurity nitrogen and a neighbor vacancy.
A Gauss magnetic field was used to split the electron energy level scheme of the NV defect creating a spin-1 system. Compatible measurements were performed by pulse sequences.
A violation of the inequality by up to $2.221$ was achieved for the neutrally polarized state identified with the five fold symmetry axis of the KCBS pentagram.
  The extremely high experimentally measured violation (2.221 is only slightly below the quantum bound of $\sqrt{5}$) is partially due to the imperfect implementation of the no-disturbance condition, symptomatic of a experimentally detected nonzero overlap between orthogonal projectors  of $\sum_{k = 1\dots 5}|\langle v_k|v_{k+1 \, {\rm mod}\, 5}\rangle|^2/5=0.0020$ on average.

Already these ideas have been transferred into the search for applications of experimental quantum contextuality. D.-L. Deng et al. demonstrated that violation of the KCBS inequality can be used to certify a lower bound on the degree of randomness of measurement outcomes for a single qutrit \cite{DZCHYWD13}. A single $800~nm$  photon which was conditionally obtained by a spontaneous parametric down conversion process and then put into a superposition of three different paths using half-wave plates and polarization beam splitters.
The purpose of the experiment was to demonstrate violation of the KCBS's inequality guaranteed a certain confidence level of randomness \cite{DZCHYWD13}.
Moreover, it was suggested that the bound on genuine randomness is given by the level of violation of the KCBS inequality.
The randomness of the output string was quantified by the min-entropy.
This implementation could generate approximately $5246$ net random bits with $99.9\%$ confidence level.
The level of violation of the inequality  obtained was within the interval $3.9$ to $3.944$.

Concurrent work by E. Amselem et al.  experimentally demonstrated  violation of Inequality (\ref{eq:Cabelloexperiment})  using a four level system encoded in the polarization and spatial path of a single photon as outlined in Equation (\ref{eq:encodinghilbertspace}) \cite{ADLPBC12}.
The $780$ {\it nm} single-photon source  was emitted by a diode laser and measured by free space displaced Sagnac interferometers. The visibility of the interferometers ranged between $90\%$ and $99\%$ allowing violation of the inequality by up to  $3.4681$ for the quantum state $\frac{1}{\sqrt{2}}(|0\rangle+|3\rangle)$. However the implementation was susceptible to the disturbed hidden variable loophole
due to the slight incompatibility of sequential measurements.

\section{Conclusions}
\label{sec:conclusions}

In the present work we have surveyed the evolution of contextuality from its conception as a purely mathematical theorem to the empirically testable phenomenon that was recently verified in laboratories. We showed that quantum contextuality is a general framework treating quantum correlations on equal footing, regardless of whether they occur between many parts of a composite system, or between properties of a single object. Finally, we also highlighted some prominent directions in contemporary research topics.

Quantum nonlocality evolved in a similar way from the early metaphysics to experimental verification in 1980's. The breakthrough for quantum nonlocality was its application to cryptography \cite{Ekert} and the invention of the quantum teleportation protocol \cite{Teleportation} which was a catalyst for the birth of quantum information theory. We believe that at the moment we are in the eve of such breakthrough for quantum contextuality.

One of the most important open problems in contextuality is its general applicability in quantum information theory. Interestingly, the special case of nonlocality is a well established resource. This nonclassical resource is intertwined with classical resources for communication. For example, it was shown that nonlocality can be simulated with communication \cite{TonerBacon}. Moreover, it is also known that nonlocality is a useful resource in communication related problems.

On the other hand, there is very little known about the relation between general contextuality and other information theoretic resources. Recently, it was proposed that contextuality for a single (local) system can be simulated in realistic theories if one uses additional memory resources \cite{Memory}. By analogy to nonlocality, it is therefore possible that contextuality might be useful in some memory related tasks, however we need to learn more about the fundamental properties of contextuality before an affirmative answer can be given.

{\bf Acknowledgements.} This work is supported by the National Research Foundation and Ministry of Education in Singapore. PK is also supported by the Foundation for Polish Science.


\begin{thebibliography}{99}


\bibitem{Bell}
J. S. Bell,
%On the problem of Hidden varibales in Quantum Mechanics
Rev. Mod. Phys. {\bf 38}, 447 (1966)

\bibitem{KS}
S. Kochen and E.P. Specker,
%The Problem of Hidden Variables in Quantum Mechanics
J. Math. Mech. {\bf 17}, 59 (1967).

\bibitem{Klyachko}
%Simple Test for Hidden Variables in Spin-1 Systems
A. A. Klyachko, M. A. Can, S. Binicioglu, and A. S. Shumovsky,
Phys. Rev. Lett. {\bf 101}, 020403 (2008).

\bibitem{Haag}
R. Haag,
{\it Local Quantum Physics: Fields, Particles, Algebras},
Springer (1996).

\bibitem{Ekert}
A. Ekert,
Phys. Rev. Lett. {\bf 67}, 661 (1991).

\bibitem{SCARANI}
A. Acin, N. Brunner, N. Gisin, S. Massar, S. Pironio, V. Scarani,
Phys. Rev. Lett. {\bf 98}, 230501 (2007).

\bibitem{ACIN}
S. Pironio, A. Acin, S. Massar, A. Boyer de la Giroday, D. N. Matsukevich, P. Maunz, S. Olmschenk, D. Hayes, L. Luo, T. A. Manning, C. Monroe,
Nature {\bf 464}, 1021 (2010).

\bibitem{EPR}
A. Einstein, B. Podolsky, and N. Rosen,
Phys. Rev. {\bf 47}, 777 (1935).

\bibitem{Aspect}
A. Aspect, J. Dalibard, and G. Roger,
Phys. Rev. Lett. {\bf 49}, 1804 (1982).

\bibitem{PM3}
N. D. Mermin,
Rev. Mod. Phys. {\bf 65}, 803 (1993).

\bibitem{Peresbook}
A. Peres,
{\it Quantum Theory: Concepts and Methods} (Dordrecht: Kluwer), (1993).

\bibitem{Garcia-Alcaine}
A. Cabello and G. Garicia-Alcaine,
%Bell-Kochen Specker theorem for any finite dimension $n\ge3$
J. Phys. A: Math. Gen. {\bf 29}, 1025â€“1036 (1996)

\bibitem{Peres}
A. Peres,
%Two simple proofs of the Kochen-Specker theorem
J. Phys. A: Math. Gen. {\bf 24} L175-L178 (1991)

\bibitem{PM1}
A. Peres,
Phys. Lett. A {\bf 151}, 107 (1990).

\bibitem{Mermin1}
N.D. Mermin,
%What's wrong with these elements of reality?
Physics Today, {\bf 43}, 9â€“11 (1990)

\bibitem{Mermin2}
N.D. Mermin,
%Simple unified form for the major no-hidden-variables theorems
Phys. Rev. Lett. {\bf 65}, 3373 (1990)

\bibitem{Aravind}
P. K. Aravind,
%QUANTUM MYSTERIES REVISITED AGAIN
Am. J. Phys. 72, 1303-7 (2004).

\bibitem{Fine}
A. Fine,
%Hidden Vaiables, Joint Probability, and the Bell Inequalities
Phys. Rev. Lett. {\bf 48}, 291 (1982).

\bibitem{LSW}
Y.-C. Liang, R. W. Spekkens, and H. M. Wiseman,
Phys. Rep. {\bf 506}, 1 (2011).

\bibitem{Ramanathan}
P. Kurzynski, R. Ramanathan and D. Kaszlikowski,
%Entropic test of quantum contextuality
Phys. Rev. Lett. {\bf 109}, 020404 (2012)

\bibitem{Clauser}
J. F. Clauser, M. A. Horne,  A. Shimony and R. A. Holt,
%Proposed Experimnet to Test Local Hidden-Variable Theories
Phys. Rev. Lett., {\bf 23}, 880 (1969).

\bibitem{ncycles}
M. Araujo, M. Tulio Quintino, C. Budroni, M. Terra Cunha, and A. Cabello,
arXiv:1206.3212 (2012).

\bibitem{CK}
J. Conway and S. Kochen,
Found. Phys. {\bf 36}, 10 (2006).

\bibitem{Gleason}
A. Gleason,
J. Math. Mech. {\bf 6}, 885 (1957).

\bibitem{CSW}
A. Cabello, S. Severini, and A. Winter,
arXiv:1010.2163 (2010).

\bibitem{Monogamy2}
R. Ramanathan, A. Soeda, P. Kurzynski, and D. Kaszlikowski,
Phys. Rev. Lett. {\bf 109}, 050404 (2012).

\bibitem{ICC}
A. Cabello,
Phys. Rev. Lett. {\bf 110}, 060402 (2013).

\bibitem{ICC2}
B. Yan,
arXiv:1303.4357 (2013).

\bibitem{AACB13}
J. Ahrens, E. Amselem, A. Cabello, and M. Bourennane,
arXiv:1301.2887
%Two Fundamental Experimental Tests of Nonclassicality with Qutrits

\bibitem{Wright}
R. Wright,
{\it Mathematical Foundations of Quantum Mechanics},
edited by A. R. Marlow (Academic Press, San Diego, 1978), p. 255.

\bibitem{Kurzynski}
% "Contextuality of almost all qutrit states can be revealed with nine observables"
P. Kurzynski, D. Kaszlikowski,
Phys. Rev. A {\bf 86}, 042125 (2012).


\bibitem{Miszczak}
J.A. Miszczak,
%Singular values decomposition and matrix reorderings in quantum information theory,
Int. J. Mod. Phys. C, {\bf 22}, 897-918 (2011)

\bibitem{Thompson}
J. Thompson, R. Pisarczyk, P. Kurzy\'nski and D. Kaszlikowski,
arXiv:1301.3496 (2013).
%An experimental proposal for revealing contextuality in almost all qutrit states

\bibitem{ADLPBC12}
E. Amselem, L.E.  Danielsen, A.J. L\'opez-Tarrida, J.R. Portillo, M. Bourennane, and A. Cabello,
Phys. Rev. Lett. {\bf 108}, 200405 (2012).
%Experimental fully contextual correlations

\bibitem{CGA1998}
A. Cabello and G. Garc\'ia-Alcaine,
Phys. Rev. Lett. \textbf{80}, 1797 (1998).

\bibitem{Meyer1999}
D.A. Meyer,
Phys. Rev. Lett. \textbf{83}, 3751 (1999).

\bibitem{Kent1999}
A. Kent,
Phys. Rev. Lett. \textbf{83}, 3755 (1999).

\bibitem{CK2000}
R. Clifton and A. Kent,
Proc. R. Soc. London, Ser. A \textbf{456}, 2101 (2000

\bibitem{SZWZ2000}
C. Simon, M. \.Zukowski, H. Weinfurter, and A. Zeilinger,
Phys. Rev. Lett. \textbf{85}, 1783 (2000).

\bibitem{HKSS2001}
H. Havlicek, G. Krenn, J. Summhammer, and K. Svozil,
J. Phys. A \textbf{34}, 3071 (2001).

\bibitem{Appleby2002}
D.M. Appleby,
Phys. Rev. A \textbf{65}, 022105 (2002).

\bibitem{SBZ2001}
C. Simon, \v{C}. Brukner, and A. Zelinger,
Phys. Rev. Lett. \textbf{86}, 4427 (2001).

\bibitem{Larsson2002}
J. A. Larsson,
Europhys. Lett. \textbf{58}, 799 (2002).

\bibitem{Cabello2002}
A. Cabello,
Phys. Rev. A \textbf{65}, 052101 (2002).

\bibitem{HLZPG2003}
Y.-F. Huang, C.-F. Li, Y.-S. Zhang, J.-W. Pan, and G.-C. Guo,
Phys. Rev. Lett. \textbf{90}, 250401 (2003).

\bibitem{Cabello}
A. Cabello,
%Experimentally Testable State-Independent Quantum Contextuality
Phys. Rev. Lett. {\bf 101}, 210401 (2008)

\bibitem{Bengtsson}
I. Benstsson, K. Blanchfield, and A. Cabello,
Phys. Lett. A \textbf{376} 374 (2010).

\bibitem{Yu2}
S. Yu and C.H. Oh,
Phys. Rev. Lett. \textbf{108}, 030402 (2012).

\bibitem{Cabello2}
A. Cabello,
arXiv:1112.5149.

\bibitem{Yu1}
S. Yu and C.H. Oh,
arXiv:1112.5513.

\bibitem{Kleinmann}
M. Kleinmann, C. Budroni, J.-A. Larsson, O. G\"uhne, and A. Cabello,
Phys. Rev. Lett. \textbf{109}, 250402 (2012).

\bibitem{ZWDCLHYD12}
C. Zu, Y. -X. Wang, D.-L. Deng, X.-Y. Chang, K. Liu, P.-Y. Hou, H.-X. Yang, L.-M. Duan,
Phys. Rev. Lett. {\bf 109}, 150401 (2012).
%State-independent experimental test of quantum contextuality in an indivisible system

\bibitem{Plastino}
\'A.R. Plastino and A. Cabello,
 Phys. Rev. A\textbf{82}, 022114 (2010).

\bibitem{Braunstein}
S. L. Braunstien and C. M. Caves,
%Information-Theoretic Bell inequalities
Phys. Rev. Lett. {\bf 61}, 662 (1988)

\bibitem{Cerf}
N. J. Cerf and C. Adami,
Phys. Rev. Lett. {\bf 79}, 5194 (1997)

\bibitem{Fritz}
R. Chaves and T. Fritz,
Phys. Rev. A {\bf 85}, 032113 (2012).

\bibitem{Monogamy1}
M. Pawlowski and C. Brukner,
Phys. Rev. Lett. {\bf 102}, 030403 (2009).


\bibitem{KZGKGCBR09}
G. Kirchmair, F. Z\"ahringer, R. Gerrritsma, M. Kleinmann, O. G\"uhne, A. Cabello, R. Blatt, and C.F. Roos,
Nature (London) {\bf 460}, 494 (2009).
%State-independent experimental test of quantum contextuality


\bibitem{ARBC09}
E. Amselem, M. Radmark, M. Bourennane, and A. Cabello,
Phys. Rev. Lett. {\bf 103}, 160405 (2009).
%State-independent quantum contextually with single photons


\bibitem{MRCL10}
O. Moussa, C.A. Ryan, D.G. Cory, and R. Laflamme,
Phys. Rev. Lett. {\bf 104}, 160501 (2010).
%Testing contextually on quantum ensembles with one clean qubit


\bibitem{LHGSZLG09}
B.H. Liu, Y.F. Huang, Y.X. Gong, F.W. Sun, Y.S. Zhang, C.F. Li, and G.C. Guo,
Phys. Rev. A {\bf 80}, 044101 (2009).
%Experimental demonstration of quantum contextuality with nonentangled photons


\bibitem{BKSSCRH09}
H. Bartosik, J. Klepp, C. Schmitzer, S. Sponar, A. Cabello, H. Rauch, and Y. Hasegawa,
Phys. Rev. Lett. {\bf 103}, 040403 (2009).
%Experimental test of quantum contextuality in neutron interferometry


\bibitem{Cabello13}
A. Cabello,
eprint arXiv:1201.0374


\bibitem{ZUZAWDSDK12}
X. Zhang, M. Um, J. Zhang, S. An, Y. Wang, D. Deng, C. Shen, L. Duan, and K. Kim,
Phys. Rev. Lett. {\bf 110}, 070401 (2013).
%State-independent experimental tests of quantum contextuality in a three dimensional system


\bibitem{DHANBSC13}
V. D'Ambrosio, I. Herbauts, E. Amselem, E. Nagali, M. Bourennane, F. Sciarrino, and A. Cabello,
Phys. Rev. X {\bf 3}, 011012 (2013).
%Experimental Implementation of a Kochen-Specker set of quantum tests


\bibitem{LLSLRWZ11}
R. Lapkiewicz, P. Li, C. Schaeff, N.K. Langford, S. Ramelow, M. Wie\'sniak, and A. Zeilinger,
Nature {\bf 474}, 490 (2011).
%Experimental non-classicality of an indivisible quantum system



\bibitem{KSSWHZJDYD12}
Xi Kong, M. Shi, F. Shi, P. Wang, P. Huang, Qi Zhang, C. Ju, C. Duan, S. Yu, and J. Du,
arXiv:1210.0961
%An experimental test of the non-classicality of quantum mechanics using an unmovable and indivisible system



\bibitem{DZCHYWD13}
D.-L. Deng, C. Zu, X.-Y. Chang, P.-Y. Hou, H.-X. Yang, Y. -X. Wang, and L.-M. Duan,
arXiv:1301.5364
%Exploring quantum contextuality to generate true random numbers



\bibitem{Teleportation}
C. H. Bennett, G. Brassard, C. Crepeau, R. Jozsa, A. Peres, and W. K. Wootters,
Phys. Rev. Lett. \textbf{70}, 1895 (1993).



\bibitem{TonerBacon}
B. F. Toner and D. Bacon,
Phys. Rev. Lett. \textbf{91}, 187904 (2003).

\bibitem{Memory}
M. Kleinmann, O. Guhne, J. R. Portillo, J.-A. Larsson, and A. Cabello
New J. Phys. {\bf 13}, 113011 (2011).





%Here ends the order in the references








%\bibitem{Clauser1}
%J. F. Clauser and A. Shimony,
%Bell's theorem: experimental tests and implications,
%Reports on Progress in Physics {\bf 41}, 1881 (1978)

%\bibitem{Clifton}
%R. Clifton,
%Getting Contextual and nonlocal elements-of-reality the easy way
%Am. J. Phys. {\bf 61}, 443 (1993)


%\bibitem{Estebaranz}
%A. Cabello, J. M. Estebaranz and G. Garcia-Alcaine,
%Bell-Kochen Specker theorem: A proof with 18 vectors
%Phys. Lett. A, {\bf 212}, 183-187 (1996)


%\bibitem{Pan}
%A.K. Pan and D. Home,
%Eur. Phys. Journal D \textbf{66}, 62 (2012).











\end{thebibliography}
\end{document}